\documentclass[hyper]{JHEP} % 10pt is ignored!

\usepackage{epsfig}

\newcommand\fverb{\setbox\pippobox=\hbox\bgroup\verb}
\newcommand\fverbdo{\egroup\medskip\noindent%
			\fbox{\unhbox\pippobox}\ }
\newcommand\fverbit{\egroup\item[\fbox{\unhbox\pippobox}]}
\newbox\pippobox
\title{D-branes in Type IIA and Type IIB theories from tachyon condensation}

\author{by J. Kluso\v{n}\\
	 Department of Theoretical Physics and Astrophysics\\
                   Faculty of Science, Masaryk University\\
Kotl\'{a}\v{r}sk\'{a} 2, 611 37, Brno\\
Czech Republic\\
	E-mail: \email{klu@physics.muni.cz}}

\preprint{\hepth{0001123}}	% OR: \preprint{Aaaa/Mm/Yy\\Aaa-aa/Nnnnnn}
\abstract{In this paper we will construct all D-branes in 
Type IIA and Type IIB theories via tachyon condensation.  Then we propose
form of Wess-Zumino term for non-BPS D-brane and we will show that tachyon 
condensation in this term leads to standard Wess-Zumino term for BPS D-brane.}


\def\tr{\mathrm{Tr}}

\begin{document}

\maketitle
\section{Introduction}
Non-BPS D-branes have been intensively studied in recent
years (see, for example \cite{Sen1,Sen2,Sen3,Sen4,Sen5}
and for review, see \cite{SenA,Lerda,Schwarz}). It was
proposed in ref. \cite{witen,Horava,Olsen} that all D-branes in Type
IIA, IIB and Type I theory can be classified via K-theory. This
classification is based on  tachyon condensation in 
unstable system of space-time filling branes, D9-branes
and antibranes in Type IIB theory and non-BPS D9-branes in 
Type IIA theory. In this approach, existence of stable BPS D-branes
was deduced from  topological arguments.
It would be nice to see how these branes emerge directly
from non-BPS D9-branes in IIA theory or from system D9-branes
and antibranes in IIB theory. In this paper we would like
to show this phenomena.

In the previous paper \cite{Kluson}, we have proposed
action for system of non-BPS D9-branes, following 
\cite{Sen}. We have shown that via tachyon condensation
in form of kink solution on the system of  $N$ non-BPS D9-branes
we are able to obtain action for $N-k$ D8-branes and
$k$ D8-antibranes. Than we have  shown that  
we are able to obtain
BPS D6-brane  from two
D9-branes in Type IIA theory in "step by step " construction, 
that is based on tachyon condensation in form of kink solution.
 We have
also shown, that the action for D6-brane 
turns out directly from action for system D8-brane and antibrane from
tachyon condensation in the form of vortex solution on 
world-volume of system brane and antibrane. We have finished
the paper \cite{Kluson} with analysing of Wess-Zumino (WZ) term
for non-BPS D-brane 
and we have discussed some problems related to
 tachyon condensation in WZ term.
in this term.

In this paper we will continue in our previous work
of tachyon condensation. The starting point will be
action for 16 non-BPS D9-branes in Type IIA theory.
The action for non-BPS D-brane was proposed in 
\cite{Sen} and we have generalised this action for
the system of $N$ non-BPS D-branes in Type IIA theory
in \cite{Kluson} in the same way as for ordinary BPS
D-brane, see for example \cite{witen3,taylor}. As was
explained in \cite{Sen}, action for non-BPS D-brane
contains term, which expresses  presence of tachyon on
the world-volume of non-BPS D-brane. This term has a property,
which lies in heart of our construction, that for tachyon
equal to its vacuum value, it is zero. We have made
some comments about this term in \cite{Kluson}, where
we have estimated its form on general grounds. In this paper
we will see, that with using this simple term, we are able
to get some interesting results. 

Plan of this paper is follows. In section (\ref{two}) we will show
how our idea works on rather simple example of "step by step"
construction of tachyon condensation on world-volume of
16 non-BPS D9-branes in Type IIA theory. We will see, that
in this construction we obtain action for 16 D0-branes in Type 
IIA theory, with agreement with \cite{Horava}, but there
is a slight difference with \cite{Horava}, where was argued
that due to the tachyon condensation we are able to get
one single D0-brane. In fact, as we will see on many examples
of tachyon condensation on world-volume of D9-branes in IIA 
theory, we always get action for 16 D-branes, some of them
form a bound state, so they do not contribute to the dynamic of
the system, but their presence can be deduced from tension of
resulting brane (for example, action that arises from tachyon
condensation in section (\ref{two}) contains factor $2^4$ expressing
the fact that  D0-brane is a bound state of 16 D0-branes).
This result is consequence of the fact, that 16 non-BPS D9-branes
participate in the construction of D-branes in Type IIA theory.

In section (\ref{three}) we will discuss construction
of general  Dp-branes in Type IIA theory via tachyon
condensation in generalised vortex solution, following
the approach in \cite{Horava}. Again we obtain correct
action for D-branes. In the end of these section, we will
show that we are also able to obtain all lower dimensional
D-branes in Type IIA theory from system of 16 non-BPS D9-branes
which is in agreement with \cite{Horava}. Again we will see
that the resulting action describes 16 D-branes.

In section (\ref{four}) we turn to the problem of construction
of non-BPS D-branes in Type IIA theory, following \cite{Horava3}.
We will show that with tachyon solution presented in \cite{Horava3}
we are able to obtain action for non-BPS D-brane in Type IIA
theory, either with direct construction presented in section (\ref{three})
or with step by step construction presented in section (\ref{two}).

In section (\ref{five}) we will discuss the construction of
BPS and non-BPS D-branes in Type IIB theory. 

In section (\ref{six}) we propose form of Wess-Zumino
term for non-BPS D-branes. We start from the WZ term
for single non-BPS D-brane presented in the ref.\cite{Billo}
and generalise this result for system of N D-branes and we also
propose higher terms in covariant derivative of tachyon, which
are needed for correct reproduction of WZ term for ordinary
BPS D-brane, as we will see on the example of tachyon
condensation in "step by step" construction leading to the D0-brane. Again
we will see that resulting charge of D0-brane contains factor
$16$. The emergence of this factor in WZ term is crucial, because
the resulting D-brane should be stable object \cite{witen,Horava} and
such a object should be BPS state of theory and consequently
charge and tension of this object must be equal.

Then we will show that tachyon condensation
in the form presented in (\ref{three}) gives the correct value
of Wess-Zumino term for BPS D-branes and non-BPS D-branes.

In section (\ref{eight}) we sum up our result and propose other
possibilities of our research.

%%%%%%%%%%%%%%%%%%%%%%%%%%%%%%%%%%%%%%%%%%%%
%%%%%Step by step %%%%%%%%%%%%%%%%%%%%%%%%%%%%%
%%%%%%%%%%%%%%%%%%%%%%%%%%%%%%%%%%%%%%%%%%%%
\section{Step by step construction}\label{two}
We would like to show, that in our approach we are able to obtain all D-branes in 
Type IIA, IIB theories. We will start with IIA theory and we will construct
D-branes with using tachyon condensation as in \cite{Horava}. Firstly,
we will show "step by step" construction, where we will construct D-branes
from kink tachyon solution. 

Our starting point is the low energy  action for $N$ non-BPS D9-branes:
\begin{eqnarray}\label{act2x}
S=-\int d^{10}x\left\{1+\frac{(2\pi\alpha')^2}{4} (\tr F_{MN}F^{MN}+2i \tr \theta_L\Gamma^MD_M\theta_L+\right. \nonumber \\
\left. +2i \tr\theta_R\Gamma^MD_M\theta_R)\right\}F(T,DT,...) \nonumber \\
\end{eqnarray}
In this section we consider only leading order terms in expansion of
DBI action, because there are some problems in generalisation of
DBI action for non-Abelian case (for review, see \cite{Tseytlin}).
 In this action: $M,N=0,...9, \ \theta_R$
is  right handed Majorana-Weyl spinor, $\theta_L$ is left handed Majorana-Weyl
spinor and $F$ is a function expressing interaction between massless fields coming
from open string sector and tachyon as well as interaction between tachyon and
fields coming from closed string sector, graviton, antisymmetric two form ...). In
this article, we are interested only in trivial background, so that there are no interaction between
tachyon and fields coming from closed string sector. As was argued in paper
\cite{Kluson}, this term has a  form
\begin{equation}\label{F2}
F(T,DT..)=\frac{\sqrt{2}2\pi}{Ng(4\pi^2\alpha')^{\frac{11}{2}}}\left [\tr D_MTD^MT+
\tr (f(T)\overline{\theta_R} \theta_L) +V(T)\right]
\end{equation}
and tachyon potential has a form \cite{Horava}:
\begin{equation}\label{pot}
V(T)=-m^2 \tr T^2+ \lambda \tr T^4+\lambda \tr T_v^4
\end{equation}
where 
\begin{equation}\label{min}
T_v^2=\frac{m^2}{2\lambda}
\end{equation}
is minimum of potential and we have included
constant term into potential  in order to have $V(T=T_0)=0$. In (\ref{F2})
we have included normalisation factor $\frac{1}{N}$ for reason, which will
be clear later. We must mention that this potential is zeroth order approximation
of potential given in \cite{Berko,SenT}. We take this simple form of potential,
because we can get analytic solution of equation of motion for tachyon.
 Finally we have included in (\ref{F2}) factor $\frac{\sqrt{2}2\pi}{g
(4\pi^2\alpha')^{\frac{11}{2}}}$, where
$g$ is a string coupling constant. This factor corresponds to the tension 
of non-BPS D9-brane.

We will demonstrate that with using this action, we are able to obtain action for
single D0-brane in Type IIA theory. As was proposed in \cite{Horava}, natural
gauge group on non-BPS D9-branes is $U(16)$. Now we show "step by step"
construction, where in each step we use tachyon kink solution.
\begin{description}
\item[Step 1] 
After variation of action (\ref{act2}), we get equation of motion of motion for tachyon fields 
\begin{equation}
\frac{\delta F}{\delta T}_{ij} -D_M
\left(\frac{\delta F}{\delta D_M T}\right)_{ij}-\partial_M(G)(D^MT)_{ij}=0
\end{equation}
where generally $ i,j=0,...,16$ and where the symbol $G$ means the first bracket in
(\ref{act2}), which includes kinetic terms for all massless fields.  We take tachyon   solution in 
the form:
\begin{equation}\label{sol}
T(x)=\left(\begin{array}{cc} T_0(x^9) 1_{8\times 8} & T(y)\delta(x^9) \\
                                    \overline{T(y)}\delta(x^9) & -T_0(x^9)1_{8\times 8} \\
\end{array}\right)
\end{equation}
where $y$ means coordinates $x^i, i=0,...,8$ and delta function in the off-diagonal 
elements has a formal meaning, which expresses the fact that off-diagonal modes are localised in
the core of the vortex. We have  also taken $T_0$ in the form
of kink solution of equation:
\begin{equation}\label{solt}
-2\frac{d^2}{d^2x}T_0(x)+\frac{d}{dT}\mathcal{V}(T)=0
\end{equation}
where $\mathcal{V}=-m^2T^2+\lambda T^4 $. We will see that the tachyon kink solution
is nonzero in region of size of string scale, so that in the zero slope limit $\alpha'\rightarrow 0$
reduces to the solution localised in single point $x=0$. This solution can serve as a justification
of approach in \cite{Kluson}, where we have taken solution in the form of step function, which
appears as a zero slope limit of ordinary kink solution.  

 Solution
of previous equation is ordinary kink solution, which can be found in many books about extended
field configurations. We will review the basic facts about this solution.

When we multiply equation (\ref{solt}) with $T'$ we
get
\begin{equation}
2T''T'=\frac{d\mathcal{V}}{dT}T'
\Rightarrow (T')^2=\mathcal{V}
\end{equation}
when we have made integration over $x$. From previous
equation we get expression
\begin{equation}
\frac{dT}{\sqrt{\mathcal{V}}}=dx
\end{equation}
and integration we get (we take condition
that for $x_0=0, T_v=0 $):
\begin{equation}
x=\int \frac{dT}{\sqrt{\lambda}
(T^2-T_v^2)}=
\frac{1}{\sqrt{\lambda}\sqrt{\frac{m^2}{2\lambda}}}
arctangh\left(\frac{T}{T_v}\right)
\end{equation}
where $T_v^2=\frac{m^2}{2\lambda}$. From previous 
equation we get
\begin{equation}
T=T_v \tanh\left(\frac{mx}{\sqrt{2}}\right)
\end{equation}
Its first derivative is equal to 
\[ T_v\frac{m}{\sqrt{2}}(1-\tanh^2\left(\frac{mx}{\sqrt{2}}\right))\]
when we put previous results into form of $F$ function
we get
\begin{equation}
F=\frac{N2\pi\sqrt{2}}{N(4\pi^2\alpha')^{\frac{11}{2}}g}\left[
\frac{T^2_0m^2}{2}\left(1-\tanh^2\left(\frac{xm}{\sqrt{2}}\right)\right)^2
+\lambda T^4_v(\tanh^2\left(\frac{mx}{\sqrt{2}}\right)-1)^2\right]
\end{equation}
or  equivalently
\footnote{In these expressions we do not include the contributions from
off-diagonal modes of tachyon. The general expression of function $F$  will
be given below.}
\begin{equation}
F=\frac{N2\pi \sqrt{2}m^4}{N(4\pi^2\alpha')^{\frac{11}{2}}g
2\lambda}(1-\tanh^2\left(\frac{mx}{\sqrt{2}}\right))^2
\end{equation}

In previous equations we have denoted $N=16$. From the behaviour 
of function $\tanh(x)$ we know
that is equal to one almost everywhere except small region which is equal
to $ (-1,1)$. From this fact we see that $F$ is zero outside the region of size of
string length $l_s$. In zero slope limit $\alpha'\rightarrow 0 \Rightarrow
m=\frac{1}{\alpha'}\rightarrow \infty $ and we see that resulting vortex is localised in
the point $x=0$. In other words, in this limit the $F$ function effectively looks like
a delta function. From this reason we have off-diagonal modes of tachyon fields localised
in the core of the vortex, which explain the presence of delta function in (\ref{sol}). In
fact, this delta function has only symbolic meaning, in calculations we will replace this
delta function by  factor $2(1-\tanh^2(\frac{mx}{2\sqrt{2}}))^2$ that can serve as 
a regularisation of  the delta function as we have seen above. 

The next calculation is the same as in \cite{Kluson} and we refer to this paper for more details. After
some calculations we get following form of $F$:
\begin{eqnarray}\label{F3}
F=\frac{2\pi \sqrt{2}}{16(4\pi^2\alpha')^{\frac{11}{2}}g
}2(1-\tanh^2\left(\frac{mx}{\sqrt{2}}\right))^2\left[
\frac{16m^4}{4\lambda}+2\tr(XT-TX')(X'\overline{T}-\overline{T}X)+\right.\nonumber \\
\left.+2\tr (\tilde{D}T^{\mu}\overline{\tilde{D}_{\mu}T}+(-m^2\tr(T\overline{T})+\lambda \tr(T\overline{T})^2)
 +\tr f(T)\overline{B}\theta)+\tr (f(\overline{T})\overline{C}\theta')\right] \nonumber \\
\end{eqnarray}
After putting (\ref{F3}) into (\ref{act2x}) we can easily make integration over $x$ due to
the fact that first bracket is independent on $x$ coordinates. This important fact
comes from the third term in (\ref{solt}), because covariant derivative with respect $x$ is nonzero
so that derivation of $G$ with respect to $x$ should be zero, so we get condition that all
massless fields are independent on $x$ coordinates.
Integration over $x$ gives

\begin{equation}
 2\int_{-\infty}^{\infty}dx
\left(1-\tanh^2(\frac{mx}{\sqrt{2}})\right)^2
=\frac{8\sqrt{2}}{3m}=\frac{8\sqrt{2}}{6\pi}
(4\pi^2\alpha')^{\frac{1}{2}}=0.606(4\pi^2\alpha')^
{\frac{1}{2}}
\end{equation}
We will discuss this numerical value in the end of the section. For the time
being we can claim the due to the tachyon condensation in the form of kink
solution we have obtained the action for 8 D8-branes and 8 D8-antibranes
\cite{Kluson}: 
\begin{eqnarray}\label{action2}
S=-0.606(4\pi^2\alpha')^{1/2}C_9\int_{R^{1,8}}d^{9}x\left[1+\frac{(2\pi\alpha')^2}{4}\left\{\tr\left( F_{\mu\nu}F^{\mu\nu}+2D_{\mu}XD^{\mu}X
+\right.\right.\right. \nonumber \\
\left.\left.\left.+2i\overline{\theta}(\Gamma^{\mu}D_{\mu}\theta+\Gamma^9[X,\theta]\right)
+4i\tr(\overline{B}\Gamma^{\mu}\tilde{D}_{\mu}B+\overline{B}\Gamma^9(XB-BX')) \right.\right. + \nonumber \\
\left.\left.+\tr\left( F'_{\mu\nu}F'^{\mu\nu}+2D'_{\mu}X'D'^{\mu}X'
+2i\overline{\theta}'(\Gamma^{\mu}D'_{\mu}\theta'+
\Gamma^9[X',\theta'])\right)\right\}\right]\times \nonumber \\
\times \frac{1}{16}\left\{16\frac{m^4}{4\lambda}+2\tr(XT-TX')(X'\overline{T}-\overline{T}X) + \right. \nonumber \\
\left.+2\tr (\tilde{D}T^{\mu}\overline{\tilde{D}_{\mu}T}+(-m^2\tr(T\overline{T})+\lambda \tr(T\overline{T})^2)\right. 
\nonumber \\
\left. +\tr f(T)\overline{B}\theta)+\tr (f(\overline{T})\overline{C}\theta')\right\} \nonumber\\
\end{eqnarray}

where $X,\theta, A$ belong to the adjoin representation of $U(8)$,which corresponds to the gauge group
of  8 D8-branes, similarly $X,\theta',A'$ correspond to 8 D8-antibranes and $T,B=C^{\dag}$ are tachyon
and spinor fields respectively coming from the string sector connecting brane and antibrane and they 
transform in the $ {\bf(8, \overline{8}) }$ of $U(8) \times U(8) $. Finally, we also define
$ DX=dX+[A,X] , \ D'X'=dX'+[A',X'] ,\tilde{DB}=dB+AB-BA' $ and we have used notation $C_p=\frac{2\pi\sqrt{2}}{
(4\pi^2\alpha')^{\frac{p+1}{2}}g}$.

 Now we proceed to the second step, which
is a tachyon condensation in the form of kink solution on the world-volume of these branes.
%%%%%%%%%%%%%%%%%%%%%%%%%%%%%%%%%%%%%%%%%%%%%%%%%%%%%%%%%%%%%%%%%
%%%%%%%%%%% S   T   E   P       2     %%%%%%%%%%%%%%%%%%%%%%%%%%%%%%%%%%%%%%%%
%%%%%%%%%%%%%%%%%%%%%%%%%%%%%%%%%%%%%%%%%%%%%%%%%%%%%%%%%%%%%%%%%

\item[Step 2]

We will construct tachyon kink solution on world-volume 8 D8-branes and 8 D8-antibranes.
The solution has a form:
\begin{equation}\label{solBaB}
T(x^8)_{ij}=T_0(x^8)\delta_{ij}
\end{equation}
where $i,j=1...8$ and where $T_0(x^8)$ is solution of (\ref{solt}). We will see that this solution place
constrains on the form of massless fields.

Firstly, we see, that (\ref{solBaB}) breaks gauge symmetry $U(8)\times U(8) $ into
diagonal subgroup $U(8)$. Now we will solve equation of motion in point $x^8\neq 0$, where
tachyon field is in its vacuum value and it is constant, so that equation of motion reduces
to the condition:
\begin{equation}
\frac{\delta F}{\delta T}=0
\end{equation}
For term with transverse fluctuation, we obtain:
\begin{equation}\label{solSC}
\frac{\delta T}{\delta T_{mn}}=(X_{ij}\delta_{mj}\delta_{kn}-\delta_{im}\delta_{jn}X'_{jk})
(X'_{kl}\overline{T}_{li}-\overline{T}_{kl}X_{li})=0
\end{equation}
We show, that solution of this equation is condition 
\begin{equation}\label{conSC}
X_{ij}=X'_{ij}
\end{equation}
When we insert (\ref{conSC}) into (\ref{solSC}) and using (\ref{solBaB}), we get:
\[(....)(X_{kl}-X'_{kl})=0 \]
In order to obtain equation of motion for tachyon, we must vary the action (\ref{action2}) with
respect the tachyon field. As in previous step we obtain the term $\partial_{x^8}(G)\tilde{D}^8T$
where $G$ means the expressions containing massless fields. Due to the fact that $\tilde{D}_8T$
is nonzero we obtain the condition that all massless fields should be independent on $x^8$. 
From this fact we can claim that previous condition (\ref{conSC}) for fields describing transverse fluctuations
hold on the whole axis $x^8$. 

In the same way we obtain condition 
\begin{equation}\label{conGF}
A^{\mu}_{ij}=A'^{\mu}_{ij}
\end{equation}
where $\mu=0,...,7$,
because kinetic term in (\ref{action2}) reduces  for constant tachyonic solution to:
\begin{equation}
(A_{\mu}T-TA'_{\mu})(A'^{\mu}\overline{T}-\overline{T}A^{\mu})
\end{equation}
With using the same arguments as for scalar fields, this condition must hold for
all $x^8$.

Now we proceed to the question of fermionic fields, for which we obtain 
following equation from varying of $F$ :
\begin{equation}\label{solFF}
\tr \left(g(T_0)\overline{B}\theta+g(T_0)\overline{C}\theta'\right)=0
\end{equation}
where $g(T_0)=\left.\frac{dg(T)}{dT}\right |_{T=T_0} $ and where we have used
the fact, that $T_0=\overline{T}_0$. We write $B=X+iY , \ C=B^{\dag}=X-iY$ where
we have defined Hermitean matrices $ X, Y$. Then (\ref{solFF}) has a form:
\begin{equation}
g(T_0)\left[(X-iY)\Gamma^0\theta+(X+iY)\Gamma^0\theta' \right]=
g(T_0)\left[ X\Gamma^0(\theta+\theta')+iY\Gamma^0(\theta-\theta')\right]=0
\end{equation}
The solution of previous equation is 
\begin{equation}\label{conFF}
X=0, \ \theta=\theta'
\end{equation}
In the same way we could take $ Y=0, \ \theta=-\theta'$, but this  result is the same 
as previous one.

We must also show that in the point $x^8$, where tachyon is in its vacuum value,
the function $F$ is zero. We know that covariant derivatives, interaction terms with fermionic fields
an scalar fields are zero. The remaining terms are
\begin{equation}
16\frac{m^4}{4\lambda}+2V(T_v)=0
\end{equation}
when we have used $16\frac{m^4}{4\lambda}=-2V(T_v)$.
When we sum up the kinetic terms for tachyon, potential terms together with constant term, we get  
 the expression:
\begin{equation}
16\frac{m^4}{2\lambda}(1-\tanh^2(\frac{mx}{\sqrt{2}}))^2
\end{equation}
which is the same expression as in the previous step, where the meaning of this formula
has been discussed. 

In previous part we have obtained number of constrain on the massless fields, that
can suggest non-BPS D7-brane. Indeed, the tachyon solution (\ref{solBaB}) is not
the most general for describing D7-brane. Remember, that in the kink solution
only real part of the tachyon field is fixed.
 We have than freedom to add to the solution (\ref{solBaB})
an imaginary part, that is function of remaining coordinates $ x^0,...,x^7$ (we denote these
coordinates as $y$ and this imaginary part is localised only in the point $x^8=0$, because
outside this point we would like to have pure vacuum). So that generalised tachyon 
field is
\begin{equation}\label{solBaB2}
T(x^8,y)=T_0(x^8)+iT(y)\delta(x^8) , \ T(y)^{\dag}=T(y)
\end{equation}
where again delta function has symbolic meaning as in previous step. 
Now we insert this tachyon field into second bracket in (\ref{action2}) and
we use the constrains for massless fields obtained in previous part, that must
hold also for generalised solution, which modifies only behaviour of tachyon field
in the core of the kink. 
Covariant derivative has  a form:
\begin{equation}
D_xT=\frac{d}{dx^8}T_0(x^8)+i(A_8T(y)-T(y)A_8)\delta(x^8)
=\frac{d}{dx^8}T_0(x^8)+i[A_8,T(y)]\delta(x^8) 
\end{equation}
where we have used $ [A,T_0]=0$. Then we obtain :
\begin{equation}
\tr D_{x^8}TD^{x^8}\overline{T}=8(\frac{d}{dx^8}T_0(x^8))^2+\tr[X^8,T]^2\delta(x^8)
\end{equation}
where $X^8=A^8, T=T(y) $.
In the action for brane+antibrane we also have  term:
\begin{equation}
\tr(XT-TX')(X'\overline{T}-\overline{T}X)
\end{equation}
and with using (\ref{conSC}, \ref{solBaB2}) this term reduces into:
\begin{equation}\label{nTT}
\tr[X^9,T(y)]^2 \delta (x^8) 
\end{equation}
Now we take $\mu=0,...7$. Then from the remaining covariant derivatives
we obtain:
\begin{equation}
D_{\mu}T(x,y)=i (\partial_{\mu}T(y)+[A_{\mu},T(y)])
\end{equation}
As a result, we obtain from kinetic term for tachyon and term containing
transverse fluctuation  (\ref{nTT}) the final   expression:
\begin{equation}\label{nKT}
8T'(x^8)^2+\tr D_{\mu}T(y)D^{\mu}T(y)+\delta_{ij}\tr[X^i,T]
[X^j,T]\delta (x^8)
\end{equation}
where $i,j=8,9$ and we have used notation $\frac{d}{dx^8}T_0=T'_0$.

Potential term has a form 
\begin{equation}\label{nPT}
V(T)=(-m^2\tr T(y)^2+\lambda \tr T(y)^4)\delta(x^8)+V(T_0)
\end{equation}
As a last step, fermionic interaction term in point $x^8=0, T_0=0$,
 reduces with using (\ref{conFF})
in 
\begin{equation}\label{nFT}
\tr\left(if(T(y))(-iY)\Gamma^0\theta-if(T(y))(iY)\Gamma^0\theta\right)\delta(x^8)=
 2\tr f(T(y))\overline{B}\theta\delta(x^8)
\end{equation}
where we have renamed $Y\rightarrow B$. In previous calculations we have used
the fact that $T(y)$ is localised in the point $x^8=0$ where $T_0(x^8)$ is equal to
zero. 
As a result, the whole second bracket in (\ref{action2}) reduces to:
\begin{eqnarray}\label{SBn}
\frac{1}{8}(8\frac{m^4}{4\lambda}+\tr D_{\mu}TD^{\mu}T+\delta_{ij}\tr [X^i,T][X^j,T]
+ \nonumber \\
+(-m^2\tr T^2+\lambda \tr T^4)+ \tr f(T)\overline{B}\theta )
\times 2(1-\tanh^2(\frac{x^8m}{\sqrt{2}}))^2 \nonumber \\
\end{eqnarray}
where we have proceed in the same way as in step one.

Now we return to the first bracket in (\ref{action2}). With using
(\ref{conSC}, \ref{conFF}, \ref{conGF}) and the fact, that all fields are
independent on $x^8$, we obtain the following results:
\begin{equation}\label{G}
\tr(F^2+F'^2) \Rightarrow \tr(2F_{\mu\nu}F^{\mu\nu} 
+4D_{\mu}X^8D^{\mu}X^8)
\end{equation}
where $D_{\mu}=\partial_{\mu}+[A_{\mu}, ]$. In the same way, we obtain
for fermionic fields ($\theta=\theta'$):
\begin{equation}\label{F1}
4i\tr \overline{\theta}(\Gamma^{\mu}D_{\mu}\theta+
\delta_{ij}\Gamma^i[X^j,\theta] )
\end{equation}
and
\begin{equation}\label{F2a}
4i\tr \overline{B}(\Gamma^{\mu}D_{\mu}B+
\delta_{ij}\Gamma^i[X^j,B])
\end{equation}
We see, that all terms have common factor $2$. 
We than obtain the final result:
\begin{eqnarray}\label{action3}
S=-2(0,606)^2C_7\int_{R^{1,7}}d^8x\left[1+\frac{(2\pi\alpha')^2}{4}\left\{\tr F_{\mu\nu}F^{\mu\nu}
+2\delta_{ij}\tr D_{\mu}X^iD^{\mu}X^j+\right.\right. \nonumber \\
\left.\left. + 2i \tr \overline{\theta}(\Gamma^{\mu}D_{\mu}\theta+
\delta_{ij}\Gamma^i[X^j,\theta])+2i \tr\overline{B}(\Gamma^{\mu}
D_{\mu}B+\delta_{ij}\Gamma^{i}[X^j,B]) \right \}\right] \times \nonumber \\
\times\frac{1}{8}\left(\tr D_{\mu}TD^{\mu}T+\delta_{ij}\tr[X^i,T][X^j,T]+
V(T)+\tr f(T)\overline{B}\theta \right) \nonumber \\ 
\end{eqnarray}
where we have included constant term $ 8\frac{m^4}{4\lambda}$ from (\ref{nKT}) into
(\ref{nPT}). This new potential has important property, that it is zero
for tachyon equal to its vacuum value.
\begin{equation}\label{nPTa}
V(T)=8\frac{m^4}{4\lambda}-\tr m^2 T^2+\tr \lambda T^4
\end{equation}
We see, that (\ref{action3}) is natural action for $8$ non-BPS D7-branes in IIA theory.
We see the factor $2$ in front of the action, which reflects the
fact that 16 D9-branes have participated in construction of 8 non-BPS
D7-branes. 

As a next step, we will construct kink solution on world-volume
of this system.

%%%%%%%%%%%%%%%%%%%%%%%%%%%%%%%%%%%%%%%%%%%%%%%%%%%%%%%%%%%%%%%%%%%%%
%%%%%%%%%%  S    T     E     P         3           %%%%%%%%%%%%%%%%%%%%%%%%%%%%%%%%%%%%%%
%%%%%%%%%%%%%%%%%%%%%%%%%%%%%%%%%%%%%%%%%%%%%%%%%%%%%%%%%%%%%%%%%%%%%
\item[Step 3]

We take kink solution in the form:
\begin{equation}\label{kink3}
T_0(x^7)=\left(\begin{array}{cc} T_0(x^7) 1_{4 \times 4} & 0 \\
                                           0         & -T_0(x^7) 1_{4 \times 4} \\ \end{array}
\right)
\end{equation}
where $T_0(x^7)$ is the solution of equation (\ref{solt}).

The discussion is the same as in step 1, so we briefly recapitulate
the result. However, there is one difference. We have  term $ \tr [X,T]^2$
 in the second bracket
in (\ref{action3}). Analysis of this term gives
the same condition as in the case of  gauge fields in step 1, namely $X$ must have a
 form :
\begin{equation}
X^i=\left(\begin{array}{cc} X^i & 0 \\
                                       0 & X'^i \\ \end{array}\right)
\end{equation}
where $X^i,X'^i \in U(4)$.

 With this kink solution, we obtain action describing 
4 D6-branes and 4 D6-antibranes, where each D6-brane or
antibrane is a bound state of two D6-branes or antibranes respectively.
The system has a gauge group $U(4)\times U(4)$. This
action is :
\begin{eqnarray}\label{action4}
S=-2(0.606)^3C_6\int_{R^{1,6}}d^7x\left[1+\frac{(2\pi\alpha')^2}{4}\left\{\tr\left( F_{\mu\nu}F^{\mu\nu}+2D_{\mu}XD^{\mu}X+
2i\overline{\theta}(\Gamma^{\mu}D_{\mu}\theta+\right.\right.\right. \nonumber \\
\left.\left.\left.+\delta_{ij}\Gamma^i[X^j,\theta]\right)
+4i\tr(\overline{B}\Gamma^{\mu}\tilde{D}_{\mu}B+\overline{B}\delta_{ij}\Gamma^i(X^jB-BX'^j)) \right.\right. + \nonumber \\
\left.\left.+\tr\left( F'_{\mu\nu}F'^{\mu\nu}+2D_{\mu}X'D'^{\mu}X'
+2i\overline{\theta}'(\Gamma^{\mu}D'_{\mu}\theta' 
+\delta_{ij}\Gamma^i[X'^j,\theta'])\right)\right\}\right]\times \nonumber \\
\times\frac{1}{8}\left\{8\frac{m^4}{4\lambda}+2\delta_{ij}\tr(X^iT-TX'^i)(X'^j\overline{T}-\overline{T}X^j) + \right. \nonumber \\
\left.+2\tr(\tilde{D}T^{\mu}\overline{\tilde{D}_{\mu}T}+V(T,\overline{T}))\right. 
\nonumber \\
\left. +\tr(f(T)\overline{B}\theta)+\tr(f(\overline{T})\overline{C}\theta')\right\} \nonumber\\
\end{eqnarray}
where $\mu, \nu =0,...,7 ; i,j=9,8,7$ and trace goes over adjoin representation
of $U(4)$ and the meaning of various fields  is the same as in
step 1. Now we proceed to the step 4.
%%%%%%%%%%%%%%%%%%%%%%%%%%%%%%%%%%%%%%%%%%%%%%%%%%%%%%%%%%%%%%%%%%%%%%%%
%%%%%%%%%%%%    S    T    E    P         4       %%%%%%%%%%%%%%%%%%%%%%%%%%%%%%%%%%%%%%%%%
%%%%%%%%%%%%%%%%%%%%%%%%%%%%%%%%%%%%%%%%%%%%%%%%%%%%%%%%%%%%%%%%%%%%%%%%
\item[Step 4]

In this step we construct kink solution on world-volume of 8 D6-branes and  8 D6-antibranes.
This kink solution is the same as in  (\ref{solBaB}), which breaks gauge symmetry $U(4)\times U(4)$
into its diagonal subgroup $U(4)$. 
As a result, we obtain action for 4 non-BPS D5 branes in IIA theory, where
each D-brane is a bound state of four D-branes:
\begin{eqnarray}\label{action5}
S=-4(0.606)^4C_5\int_{R^{1,5}}d^6x\left[1+\frac{(2\pi\alpha')^2}{4}\left\{\tr F_{\mu\nu}F^{\mu\nu}
+2\delta_{ij}\tr D_{\mu}X^iD^{\mu}X^j+\right.\right. \nonumber \\
\left.\left. + 2i \tr \overline{\theta}(\Gamma^{\mu}D_{\mu}\theta+
\delta_{ij}\Gamma^i[X^j,\theta])+2i \tr\overline{B}(\Gamma^{\mu}
D_{\mu}B+\delta_{ij}\Gamma^{i}[X^j,B]) \right \}\right] \times \nonumber \\
\times\frac{1}{4}\left(\tr D_{\mu}TD^{\mu}T+\delta_{ij}\tr[X^i,T][X^j,T]+
V(T)+\tr f(T)\overline{B}\theta \right) \nonumber \\ 
\end{eqnarray}
where $\mu,\nu=0,...,5; i, j=6,...,9$.
 Now we are going to the next step.
%%%%%%%%%%%%%%%%%%%%%%%%%%%%%%%%%%%%%%%%%%%%%%%%%%%%%%%%%%%%%%
%%%%%%%%%%%%%    S    T    E    P            5          %%%%%%%%%%%%%%%%%%%%%%%%%%%%
%%%%%%%%%%%%%%%%%%%%%%%%%%%%%%%%%%%%%%%%%%%%%%%%%%%%%%%%%%%%%%
\item[Step 5]

Again we construct kink solution on world-volume of four non-BPS 
D5-~branes and as a  result, we obtain action for two D4-branes and
two D4-antibranes (As in previous parts, each D4-brane is a bound state
of four D4-brane, and each D4-antibrane is a bound state of 
four D4-antibranes):
\begin{eqnarray}\label{action6}
S=-4(0.606)^5C_4\int_{R^{1,4}}d^5x\left[1+\frac{(2\pi\alpha')^2}{4}\left\{\tr\left( F_{\mu\nu}F^{\mu\nu}+2D_{\mu}XD^{\mu}X+
2i\overline{\theta}(\Gamma^{\mu}D_{\mu}\theta+\right.\right.\right. \nonumber \\
\left.\left.\left.+\delta_{ij}\Gamma^i[X^j,\theta]\right)
+4i\tr(\overline{B}\Gamma^{\mu}\tilde{D}_{\mu}B+\overline{B}\delta_{ij}\Gamma^i(X^jB-BX'^j)) \right.\right. + \nonumber \\
\left.\left.+\tr\left( F'_{\mu\nu}F'^{\mu\nu}+2D_{\mu}X'D'^{\mu}X'
+2i\overline{\theta}'(\Gamma^{\mu}D'_{\mu}\theta'+ 
\delta_{ij}\Gamma^i[X'^j,\theta'])\right)\right\}\right]\times \nonumber \\
\times \frac{1}{4}\left\{4\frac{m^4}{4\lambda}+2\delta_{ij}\tr(X^iT-TX'^i)(X'^j\overline{T}-\overline{T}X^j) + \right. \nonumber \\
\left.+2\tr(\tilde{D}T^{\mu}\overline{\tilde{D}_{\mu}T}+V(T,\overline{T}))\right. 
\nonumber \\
\left. +\tr(f(T)\overline{B}\theta)+\tr(f(\overline{T})\overline{C}\theta')\right\} \nonumber\\
\end{eqnarray}
where $\mu,\nu=0,...,4; i,j=5...,9$.
Now we arrive to the next step.
%%%%%%%%%%%%%%%%%%%%%%%%%%%%%%%%%%%%%%%%%%%%%%%%%%%%%%%%%%%%%%%%%
%%%%%%   S T E P    6 %%%%%%%%%%%%%%%%%%%%%%%%%%%%%%%%%%%%%%%%%%%%%%%%%
%%%%%%%%%%%%%%%%%%%%%%%%%%%%%%%%%%%%%%%%%%%%%%%%%%%%%%%%%%%%%%%%%
\item[Step 6]
We construct kink solution on world-volume of 2 D4-branes and 2D4-antibranes. Again, this
solution breaks symmetry $U(2)\times U(2)$ into diagonal subgroup $U(2)$. As a result, we obtain action
for two non-BPS D3-branes in IIA theory:
\begin{eqnarray}\label{action7}
S=-8(0.606)^6C_3\int_{R^{1,3}}d^4x\left[1+\frac{(2\pi\alpha')^2}{4}\left\{\tr F_{\mu\nu}F^{\mu\nu}
+2\delta_{ij}\tr D_{\mu}X^iD^{\mu}X^j+\right.\right. \nonumber \\
\left.\left. + 2i \tr \overline{\theta}(\Gamma^{\mu}D_{\mu}\theta+
\delta_{ij}\Gamma^i[X^j,\theta])+2i \tr\overline{B}(\Gamma^{\mu}
D_{\mu}B+\delta_{ij}\Gamma^{i}[X^j,B]) \right \}\right] \times \nonumber \\
\times \frac{1}{2}\left(\tr D_{\mu}TD^{\mu}T+\delta_{ij}\tr[X^i,T][X^j,T]+
V(T)+\tr f(T)\overline{B}\theta \right) \nonumber \\ 
\end{eqnarray}
where $\mu,\nu=0,...,5; i, j=6,...,9$. Again we must mention, that
each D3-brane is a bound state of 8 D3-branes. 
%%%%%%%%%%%%%%%%%%%%%%%%%%%%%%%%%%%%%%%%%%%%%%%%%%%%%%%%%%%%%
%%%%%%%%%%%%%  S   T    E    P      7    %%%%%%%%%%%%%%%%%%%%%%%%%%%%%%%%%%
%%%%%%%%%%%%%%%%%%%%%%%%%%%%%%%%%%%%%%%%%%%%%%%%%%%%%%%%%%%%%
\item[Step 7]

We construct kink solution on world-volume of two non-BPS D3-branes in IIA theory, which
leads to system of D2-brane and D2-antibrane.
This solution breaks gauge group $U(2)$ into
$U(1) \times U(1)$. As a result, all commutators
are zero and we must replace all
covariant derivatives with ordinary derivatives.  We than obtain action for D2-brane and D2-antibrane:
\begin{eqnarray}\label{action8}
S=-8(0.606)^7C_2\int_{R^{1,2}}d^3x\left[1+\frac{(2\pi\alpha')^2}{4}\left\{ F_{\mu\nu}F^{\mu\nu}+
2\delta_{ij}\partial_{\mu}X^i\partial^{\mu}X^j+
2i\overline{\theta}\Gamma^{\mu}\partial_{\mu}\theta+\right.\right. \nonumber \\
\left.\left.
+4i(\overline{B}\Gamma^{\mu}\tilde{D}_{\mu}B+\overline{B}\delta_{ij}\Gamma^i(X^jB-BX'^j)) \right.\right. + \nonumber \\
\left.\left.+\left( F'_{\mu\nu}F'^{\mu\nu}+2\delta_{ij}\partial_{\mu}X'^i\partial^{\mu}X'^j
+2i\overline{\theta}'\Gamma^{\mu}\partial_{\mu}\theta' 
\right) \right\}\right]\times \nonumber \\
\times \frac{1}{2} \left\{2\frac{m^4}{4\lambda}+2\delta_{ij}\tr(X^iT-TX'^i)(X'^j\overline{T}-\overline{T}X^j) + \right. \nonumber \\
\left.+2\tr(\tilde{D}T^{\mu}\overline{\tilde{D}_{\mu}T}+V(T,\overline{T}))\right. 
\left. +\tr(f(T)\overline{B}\theta)+\tr(f(\overline{T})\overline{C}\theta')\right\} \nonumber\\
\end{eqnarray}
where $\mu,\nu=0,...,2; i,j=3...,9$.

%%%%%%%%%%%%%%%%%%%%%%%%%%%%%%%%%%%%%%%%%%%%%%%%%%%%%%%%%%%%%%%
%%%%%%%%  S   T   E   P     8    %%%%%%%%%%%%%%%%%%%%%%%%%%%%%%%%%%%%%%%%%%
%%%%%%%%%%%%%%%%%%%%%%%%%%%%%%%%%%%%%%%%%%%%%%%%%%%%%%%%%%%%%%%
\item[Step 8]

 We consider kink solution on world-volume of D2-brane and D2-antibrane, which leads
to the  action for one non-BPS D1-brane:
\begin{eqnarray}\label{action9}
S=-16(0.606)^8C_1\int_{R^{1,1}}d^2x\left[1+\frac{(2\pi\alpha')^2}{4}\left\{ F_{\mu\nu}F^{\mu\nu}
+2\delta_{ij}\partial_{\mu}X^i\partial^{\mu}X^j+\right.\right. \nonumber \\
\left.\left. + 2i  \overline{\theta}\Gamma^{\mu}\partial_{\mu}\theta+
+2i \overline{B}\Gamma^{\mu}
\partial_{\mu}B \right \}\right] \times \nonumber \\
\times\left(\partial_{\mu}T\partial^{\mu}T+
V(T)+f(T)\overline{B}\theta \right) \nonumber \\ 
\end{eqnarray}
where $\mu,\nu=0,1 , i,j=2....9$ 

%%%%%%%%%%%%%%%%%%%%%%%%%%%%%%%%%%%%%%%%%%%%%%%%%%%%%%%%%
%%%%%%%%  S   T   E   P     9     %%%%%%%%%%%%%%%%%%%%%%%%%%%%%%%%%%%
%%%%%%%%%%%%%%%%%%%%%%%%%%%%%%%%%%%%%%%%%%%%%%%%%%%%%%%%%
Finally, we construct tachyon solution on world-volume of non-BPS D1-brane. Then, following
\cite{Kluson}, the second bracket reduces to the form 
\begin{equation}
2(1-\tanh^2\left(\frac{mx}{\sqrt{2}})\right)^2V(T_v)
\end{equation}
and integration of previous expressions gives
\begin{equation}
0.606 (4\pi^2\alpha')^{1/2} V(T_v)
\end{equation}
Now we use the results from \cite{Berko,SenT}, where it was shown that vacuum value
of potential is equal to $0.60$ of value of the mass of non-BPS D-brane. Since tachyon
potential $v(T)$ presented in ref.\cite{SenT} is related to our potential with rescaling
$V(T)=2v(T)$, because in normalisation of kinetic term in our action the factor $1/2$ is missing,
we can claim that vacuum value of potential s equal to 
\[ V(T_v)=1.2\frac{1}{\sqrt{2}} \]
than previous expressions leads to
\begin{equation}
(4\pi^2\alpha')^{1/2}0.606V(T_v)=(0.73)(4\pi^2\alpha')^{1/2}\frac{1}{\sqrt{2}}
\end{equation}
we see that  tension of D-brane that arises from single tachyon kink is about $0.73$ of tension for
D-brane, which is in agreement with result \cite{SenT}. Of course, for D0-brane the resulting tension
is much smaller than expected answer, which is result of our rough approximation. We can expect
on the grounds given in ref.\cite{SenT}, that further correction in the tachyon potential will lead to
correct results.  

To sum up, In the first bracket the fermion field $B$ is identically zero and we finish with
action for D0-brane in type IIA theory (more precisely, we end with action 
for 16 D0-branes in IIA theory, that form a bound state, but we will discuss this issue
later), with completely agreement with \cite{Horava}
\begin{equation}
S=-16(0,606)^91.2\frac{2\pi}{(4\pi^2\alpha')^{1.2}
g}\int dt\left[1+\frac{(2\pi\alpha')^2}{4}(2\partial_tX^i\partial_tX^i+2i\overline{\theta}
\Gamma^0\partial_t\theta )\right]
\end{equation}

We see, that with
this "step by step" construction we are able to obtain all D-branes
in IIA theory, and when we start with system D9-branes and 
D9-antibranes, following \cite{witen}, we are able to construct
all D9-branes in IIB theory as well. However, we would like to
see, whether direct construction, presented in \cite{witen,Horava}, 
can be applied in this approach. We return to this question in
next section.

\end{description}
%%%%%%%%%%%%%%%%%%%%%%%%%%%%%%%%%%%%%%%%%%%%%%%%%%%%%%%%%%%%%%
%%%%%%%%%%%   D I R E C T   M E T H O D (following Horava) %%%%%%%%%%%%%%%%%%%%
%%%%%%%%%%%%%%%%%%%%%%%%%%%%%%%%%%%%%%%%%%%%%%%%%%%%%%%%%%%%%

\section{Direct construction}\label{three}
In this section we show that we can construct lower dimensional
BPS D-brane also directly following \cite{Horava}, where was
argued that D-brane of codimension $2k+1$ can be construct
as vortex solution on world-volume of $2^k$ non-BPS D9-branes
with gauge group $U(2^k)$.
In region around point $x=0$, the tachyon field looks like:
\begin{equation}\label{solD}
T(x)=\sum_{i=1}^{2k+1}\Gamma_ix^i
\end{equation}
where $\Gamma_i$ are Gamma matrices of group $SO(2k+1)$,
which is a symmetry group of transverse space to D(8-2k)-brane
and $x^i,i=1...,2k+1 $ are coordinates on this transverse space.

It was argued \cite{Horava} that this tachyon condensation is equivalent to 
condensation of tachyon in step by step construction, where tachyon
forms a kink solution in each step. In order to obtain correct kink
solution, we generalise previous equation in the form
\begin{equation}\label{solD1}
T=\sum_{i=1}^{2k+1}\Gamma_iT_i(x^i)
\end{equation}
we will see that tachyon condensation in the form of this field is
equivalent to condensation of tachyon in form of step by step construction.
We start to solve the equation of motion for tachyon. We write
the action for non-BPS D-brane:
\begin{eqnarray}\label{act2}
S=-C_{9}\int d^{10}x\left\{1+\frac{(2\pi\alpha')^2}{4} (\tr F_{MN}F^{MN}+2i \tr \theta_L\Gamma^MD_M\theta_L+\right. \nonumber \\
\left. +2i \tr\theta_R\Gamma^MD_M\theta_R)\right\}F(T,DT,...) \nonumber \\
\end{eqnarray}
where $C_{p}=\frac{2\pi\sqrt{2}}{g(4\pi^2\alpha')^{\frac{p+1}{2}}}$
and we implicitly work in limit $\alpha'\rightarrow 0$, because
than we can neglect the higher terms in expansion of Born-Infeld
action. In previous expression the $F$ function express integration
between tachyon and other fields and has a form:
\begin{equation}\label{F2c}
F(T,DT..)=\frac{1}{2^k}\left [\tr D_MTD^MT+
\tr (f(T)\overline{\theta_R} \theta_L) +V(T)\right]
\end{equation}
and $V$ is a potential for tachyon (we again work in zeroth approximation,
which allows analytic solution \cite{SenT}).
\begin{equation}
V(T)=-\tr m^2 T^2+\lambda \tr T^4+\tr T_v^4
\end{equation}
where $T_v=\frac{m^2}{2\lambda}$ is a vacuum value of tachyon
field. 
Equation of motion for tachyon field has a form:
\begin{equation}
-\partial_{M}(G)D^{M}T+G(-2D_{M}D^{M}T
+\frac{\partial U}{\partial T})=0
\end{equation}
where $G$ means the first bracket in action for non-BPS D-brane.
We know that tachyon is a function only $2k+1$ transverse coordinates
so that around the core we have $D_iT\neq 0$ so that in order
to obey equation of motion we must pose the requirement that
all fields, which are present in $G$ should be independent on transverse
coordinates. We will see that this is natural requirement because the
effective size of the core is of order string scale and we know that
BI action is valid only for slowly varying fields so that these fields do not
change in region of size of string scale. In solving previous equation
we must also demand the vanishing covariant derivative with tangent 
direction to the vortex otherwise we should take $\partial_M G=0$ for
all $M$ and we do not get any interesting dynamical system. From the
condition of vanishing the covariant derivative $D_{\mu}T, \ \mu=0,...,8-2k$ we get 
condition on gauge field:
\begin{equation}
D_{\mu}T=[A_{\mu},T]=0 \Rightarrow A_{\mu} \in SU(2^k)=0
\end{equation}
and only $\mathcal{A}_{\mu} \in U(1)$ remains as a free dynamical field.

Suppose now that we are in the region out of the core where tachyon is
in its vacuum value. Than its derivative is zero and we get the same condition
for gauge fields as in previous case:
\begin{equation}
DT_i=0=[A_i,T]=0\Rightarrow A_i=0
\end{equation}
 and only $\mathcal{A}_i \in U(1)$ remains undetermined free dynamical field.
This condition hold almost on the whole plane and from the previous
result, which says that massless fields are not functions of transverse coordinates,
 we get condition that $A_i \in SU(2^k)$ is zero everywhere. 
With using these facts the covariant derivative for tachyon reduces to ordinary
derivative and variation of kinetic term gives (in the following we write $T_i(x^i)=T_i$:
\begin{equation}
\tr\delta(d_M(\Gamma_iT^id^M(\Gamma_jT^j)\Rightarrow
-\tr2\delta T^i \Gamma_id_Md^M(\Gamma_jT^j)=-22^k
\delta T_i d_Md^MT_i
\end{equation}
where we have used $\tr \Gamma_i\Gamma_j=2^k
\delta_{ij}$. The variation of potential term gives
\begin{eqnarray}
\delta V=\delta T^i(-2m\tr\Gamma_i\Gamma^jT_j+4\lambda
\tr\Gamma_i\Gamma^j\Gamma^k\Gamma^lT_jT_kT_l)\nonumber \\
=\delta T_i(-2m T^i+4\lambda T_i(T_jT^j) \nonumber \\
\end{eqnarray}
and variation of interaction term between fermions and tachyon gives
\begin{eqnarray}
\tr(\delta(a_1T+a_3T^3+...)\overline{\theta}_R\theta_L)=
\delta T_i \tr\Gamma^i(a_1+a_3\Gamma^i\Gamma^jT_iT_j+..)\overline{\theta_R}
\theta_L= \nonumber \\
=\delta T_i\tr\Gamma^i \frac{df(T^2)}{dT}\overline{\theta}_R\theta_L
\nonumber \\
\end{eqnarray}
where $T^2=T_iT^i$.

Now equation of motion have a form:
\begin{equation}
-2d_id^iT^i+(-2m^2T_i+4\lambda T_i^3+4\lambda T_i(\sum_{j\neq i} T_jT_j)
)+\tr (\Gamma_i \frac{df}{dT}\overline{\theta}_R\theta_L)=0 ; i=1,...,2k+1
\end{equation} 
In previous equation we do not sum over $i$ and we have used the fact 
that $d_MT_i=\delta_M^i\partial_iT(x^i)$. Since $T_i$ are  all independent,
we will see that solution of motion of these equations leads to the kink solutions
for all $T_i$ with additional conditions, which must vanish separately. Firstly,
we must pose the condition:
\begin{equation}
T_i(\sum_{j\neq i} T^jT^j)=0
\end{equation}
Outside the core of the vortex, we have $T_i \neq 0 $ so that we
must have $T_j$ to be zero. In the point $x^i=0$ we have solution
$T(x^i)=0$ so that $T(x^j)$ should be nonzero. But this is nothing
else than tachyon condensation in the form of step by step construction,
where each resulting tachyon is localised on world-volume of non-BPS
D-brane or system of branes and antibranes that arises from tachyon
condensation in the previous step. The previous condition has the
same meaning, but now we do not prefer some particular direction
where we start the tachyon condensation. In other words, this tachyon
condensation is naturally transversally invariant.
With the requirement that fermionic term should be zero separately
we get the equation for each $T_i$ in the form:
\begin{equation}
-2d_id^iT(x^i)+\frac{dU(T(x^i)}{dT}=0
\end{equation}
which has a natural solution in the form of kink solution as we have seen in the
previous section. The form of the kink solution has a form:
\begin{equation}
T(x^i)=T_v \tanh\left(\frac{mx^i}{\sqrt{2}}\right)
\end{equation}

Now
we put $T=\sum_{i=1}^{2k+1}\Gamma_iT^i$ into the $F$ function. At this
point we must be more careful, because we know that $T_i$ live only
in the point $x^j=0 , j\neq i$. We will have this fact   in the mind when we put
previous equation into $F$ function and in resulting expression we multiply
each term, that is a function of $T_i$ only, with the factor of convergence, which will
have properties of delta function and will express the above condition.
\begin{equation}
\tr d_M(\Gamma_iT^i)d^M(\Gamma_jT^j)=
2^k\sum_{i=1}^
{2k+1}\partial_iT^i\partial_iT^i
\end{equation}
and 
\begin{equation}
V(T)=-\tr m^2(T^iT^j\Gamma_i\Gamma_j)+
\lambda\tr(\Gamma_i\Gamma_j\Gamma_k\Gamma_l
T^iT^jT^kT^l)=2^k\sum_{i=1}^{2k+1}(-m^2T_i^2+
\lambda T_i^4)
\end{equation}
where we have used the fact that expression $T_iT^j , i\neq j$
is zero from arguments presented above. Then we obtain the form
of $F$ function (We use the fact that fermionic terms is zero as will
be shown in a moment):
\begin{equation}\label{FT}
F=\left[\left(\left(\frac{dT_1}{dx^1}\right)^2+(-m^2T^2_1+\lambda T^4_1)+\frac{m^4}{4\lambda}\right)
+\sum_{i=2}^{2k+1}F_i\right]
\end{equation}
where $F_i$ have a form
\begin{equation}
F_i=\left(\frac{dT_i}{dx^i}\right)^2-m^2T_i^2+\lambda T^4_i
\end{equation}
We see that we have the similar result as in the step by step construction. The first
term in (\ref{FT}) has the form
\begin{equation}
\frac{m^4}{2\lambda}(1-\tanh^2\left(\frac{mx_i}{\sqrt{2}}\right))^2
\end{equation}
Since we know that all $F_i ,i\neq 1$ are localised in the point $x^1=0$ we multiply
the second term in (\ref{FT}) with the factor $2(1-\tanh^2(\frac{mx}{\sqrt{2}}))^2$ as
in previous section. Then we can make integration over $x^1$ leading to the result
$ (0,606)(4\pi^2\alpha')^{1/2}$ in front of $F$ function and to the emergence of constant
term $V(T_v)$ in $F$ function (we have again used the fact that all fields in the first bracket
in (\ref{act2}) are independent on $x^1$). After repeating this calculation the $F$ function will give
the contribution $ (0.606)^{2k+1}1.2\frac{1}{\sqrt{2}}(4\pi^2\alpha')^{\frac{2k+1}{2}}$.

In previous part we have anticipated that fermionic terms is equal to zero. In this paragraph
we show that this is really true.
We expand the fermionic fields  in following way
\footnote{In fact, we should consider more general expansion
of $\theta $ in the form of completely antisymmetric basic. However
it is easy to see that higher terms in this expansion are identically zero.
Consider for example $\theta_{R,L}^{ij}\Gamma_{ij}$. Then we obtain two
conditions :$\tr \Gamma_i\Gamma_{kl}\overline{\theta}^{kl}_R\theta_L^0=0,
\tr \Gamma_i\Gamma_{kl}\Gamma_{mn}\overline{\theta}_R^{kl}
\theta^{mn}_L=0$. Then the second equation gives condition $\theta_L^{mn}=0$
and the first equation gives condition $\theta_R^{kl}=0$. This arguments holds
for higher terms as well.}:
\begin{equation}
\theta_{L,R}=\theta_{L.R}^0+\Gamma_i\theta^i_{L,R}
\end{equation}
and after putting this expansion into expression $\tr \Gamma^i\overline{\theta}_R
\theta_L$ we get the equations (we have used the fact that $\frac{df}{dT}$ is
nonzero. In the following we will not write this factor):
\begin{eqnarray}
\tr (\Gamma_i\Gamma_k\Gamma_l)\overline{\theta}^k_R
\theta^l_L=0 \\ \nonumber
\tr(\Gamma_i\Gamma_l)\overline{\theta}^l_R\theta_L^0 =
2^k\overline{\theta}_R^i\theta^0_L=0 \nonumber \\
2^{k} \overline{\theta}^0_R\theta^i_L=0 \nonumber \\
\end{eqnarray}
Solution of the last equation is $\overline{\theta}^0_R=0$
or $\theta^i_L=0$, but from the fact that the first equation for
$i=k=l$ gives condition $\overline{\theta}^i_R\theta^i_L=0$,
which can be solved as $\theta_L^i=0$, we see that we must
take as a solution of last equation the condition $\theta_R^0=0$
and solution of the second equation as a $\theta^i_R=0$.

 In other
words we get the result that spinor field $\theta_R$ completely disappears
as it should for restoring the BPS D-brane. We also see that only
$U(1)$ parts of $\theta_L$ remains as a free dynamical field which is in
agreement with the number of bosonic degrees of freedom. 
The vanishing
of $\theta_L \in SU(2^k)$ can be view as a consequence of vanishing
of $A\in SU(2^k)$, since both fields are related through supersymmetric
transformations of nonlinearly realised supersymmetry.

Now we can sum the results. From the fact that all fields in the adjoin representation
of $SU(2^k)$ and spinor $\theta_R$ is zero we obtain  the action for
Dp-brane of codimension $2k+1$, when we use the fact that all fields are independent
on transverse coordinates:
\begin{equation}
S=-2^k(0.606)^{2k+1}1.2T_p\int_{R^{1,8-2k}}d^{p+1}x \left[1+\frac{(2\pi\alpha')^2}{4}(F_{\mu\nu}F^{\mu\nu}
+2\partial_{\mu}X^i\partial^{\mu}X^i
+2i\overline{\theta}
\Gamma^{\mu}\partial_{\mu}\theta )\right]
\end{equation}
where $p=8-2k$ and $T_p=\frac{2\pi}{g(4\pi^2\alpha')^{\frac{p+1}{2}}}$. In previous action
we have used
\begin{equation}
F_{i\mu}F^{i\mu}=2\partial_{\mu}A_i
\partial^{\mu}A_i=2\partial_{\mu}X^i\partial^{\mu}
X^i
\end{equation}
For  D0-branes, $k=4$ and we obtain the same result as in
previous section. Again we see the presence of factor $2^k$ in front of the action, which suggests that
resulting configuration corresponds to the bound state of $2^k$ Dp-branes. We will
return to this issue in the end of this section.

 It is also important that with 16 non BPS D9-branes, we 
are able to construct lover dimensional brane as well. 
Consider general D-brane of codimension $2k+1$. As was
explained in ref.\cite{Horava}, this brane can be constructed
from $2^k$ non-BPS D9-branes with gauge group $U(2^k)$, where
tachyon $t(x) \in U(2^k)$ is a function of $2k+1$ coordinates and its form
is the same as in (\ref{solD1}). We
can put this tachyon field into $U(16)$ as follows
\begin{equation}
T(x)=t(x)\otimes \mathit{1}
\end{equation}
where $\mathit{1}$ is  $2^{4-k}\times 2^{4-k}$ unit matrix.
We can take gauge field in the form
\begin{equation}
A=\mathbf{A}*\mathrm{1}\otimes\mathit{1}+
\mathrm{A}\otimes \mathit{1}+\mathrm{1}\otimes \mathit{A}
\end{equation}
where $\mathbf{A}$ is $U(1)$ part of gauge field, $
\mathrm{A} \in SU(2^k), \mathit{A} \in SU(2^{4-k}) $ and
$\mathrm{1} $ is $2^k\times 2^k$ unit matrix. Then we can proceed
 as in previous part, because it is easy to see, that only $\mathrm{A}$
is fixed with tachyon solution (it is equal to zero) due to the fact that
\begin{equation}
[t(x)\otimes \mathit{1},\mathrm{1}\otimes \mathit{A} ]=0
\end{equation}
so that $\mathit{A}$ and $\mathbf{A}$ are
 not fixed with tachyon solution, because do not
appear in covariant derivative $DT$. Then it is also clear, that
\begin{equation}
F=\mathrm{F}\otimes \mathit{1}+\mathrm{1}\otimes \mathit{F}+
\mathbf{F} * \mathrm{1}\otimes \mathit{1}
\end{equation}
where $F$ is a field strength for $A$, $\mathrm{F}$ is a 
field strength for $\mathrm{A}$, $\mathit{F}$ is a field strength for
$\mathit{A}$ and $\mathbf{F}$ is a field strength for $\mathbf{A}$.
 Then  kinetic term $F^2$ reduces into
($\tr (A\otimes B)=\tr (A) \tr (B) $)
\begin{equation}
\tr(F^2)=\tr (\mathrm{F}^2)\tr (\mathit{1})+
2\tr \mathrm{F}\tr \mathit{F}+2\mathbf{F}(\tr \mathrm{F}\tr
\mathit{1}+\tr\mathrm{1} \tr \mathit{F})+
\tr \mathrm{1}\tr(\mathit{F}^2)+\mathbf{F}\tr\mathrm{1}\tr\mathit{1} 
\end{equation}
We can immediately see, that second and third term vanishes due to the fact, that
$\tr \mathrm{F}=\tr \mathit{F}=0 $.  When we 
combine $\mathbf{F}$ with $\mathit{F}$ into one single field
$\mathcal{F}$ in adjoin representation of $U(2^{4-k})$  and use the fact, 
that all $\mathcal{A}$ are not function of $2k+1$ coordinates,
the kinetic term reduces into
\begin{equation}
2^k(\tr \mathcal{F}_{\mu\nu}\mathcal{F}^{\mu\nu}
+2\tr D_{\mu}\mathcal{X}^iD^{\mu}\mathcal{X}^i )
\end{equation}
where $2^k$ comes from $\tr \mathrm{1} $ and $\mathcal{X}^i, i=9-2k,...,9$ are
dynamical fields describing transverse fluctuation of $2^{4-k}$ D-branes
of codimension $2k+1$ and we have also defined
$D\mathcal{X}=d\mathcal{X}+[\mathcal{A},
\mathcal{X}]$.

Analysis of fermions is the same as in case of D0-brane. Again, $\theta_R$ is 
zero and we write $\theta_L$ in the same way as $A$
\begin{equation}
\theta_L={\Theta}_L *\mathrm{1}\otimes
\mathit{1}+\tilde{\theta}_L\otimes \mathit{1}+
\mathrm{1}\otimes \phi_L 
\end{equation}
where the $\Theta_L$ is $U(1)$ part of $\theta_L$,
$\tilde{\theta}_L \in SU(2^k) $ and $\phi \in SU(2^{4-k})$.
Again $\theta_R,\tilde{\theta}_L$ are equal to zero thanks
to form of tachyon field $T \sim t\otimes 1 $. Nonzero dynamical
fields are $\Theta_L,\phi$ and as in case of gauge field
 we combine $ \Theta$
with $\phi$ into one single massless fermionic field 
$\theta \in U(2^{4-k})$, which has a kinetic and interaction
term in resulting action for D-brane
\begin{equation}
2i\tr \overline{\theta}_L\Gamma^{M}D_M\theta_L 
\Rightarrow 2^k(2i\tr \overline{\theta}\Gamma^{\mu}
D_{\mu}\theta+2i\tr \overline{\theta}\Gamma^i[\mathcal{X}^i,
\theta] )
\end{equation}
Finally, with using the fact that $F$ function gives the same
contribution as before, we obtain the action
\begin{eqnarray}\label{d}
 S=-2^k(0.606)^{2k+1}1.2T_p\int_{R^{1,8-2k}}d^{p+1}x \left[1+
\frac{(2\pi\alpha')^2}{4}(
\tr\mathcal{F}_{\mu\nu}\mathcal{F}^{\mu\nu}
+2\tr D_{\mu}\mathcal{X}^iD^{\mu}\mathcal{X}^i \right.\nonumber \\
\left.+2i\tr \overline{\theta}\Gamma^{\mu}
D_{\mu}\theta+2i\tr \overline{\theta}\Gamma^i[\mathcal{X}^i,
\theta] )\right] \nonumber \\
\end{eqnarray}
This action describes $2^{4-k}$ Dp-branes of codimension
$2k+1$, where each D-brane is a bound state of $2^k$ 
D-branes of the same codimension, which can be seen 
from the factor $2^k$ in front of the action.
These results can also be seen  in "step by step" construction.  
Consider, for example, "step by step" construction for D6-brane.
This can be schematically written as:
\begin{equation}
U(16)\stackrel{kink}{\rightarrow} U(8)\times U(8)
\stackrel{kink ,2}{\rightarrow} U(8) 
\stackrel{kink}{\rightarrow} U(8)
\end{equation}  
where factor on the second arrow expresses the presence of  factor two
in front of the action. This sequence correspond to the sequence of branes
\[ 16 D9 \rightarrow 8 D8+8\overline{D8} \rightarrow
8  D7 \rightarrow 8 D6 \]
Which implies, that this configuration describes system of 8 D6-branes with
gauge group U(8) (In fact, as was explained in previous section, each
D6-brane is a bound state of two D6-branes). The same "step by step" construction
can be used for other D-branes. For D4-brane, we have sequence:
\begin{equation}
U(16)\stackrel{kin}{\rightarrow}U(8)\times U(8) 
\stackrel{kink,2}{\rightarrow} U(8) \stackrel{kink}
{\rightarrow} U(4)\times U(4) \stackrel{kink,4}{\rightarrow}
U(4)\stackrel{kink}{\rightarrow} U(4)
\end{equation}
which corresponds to emergence of action for 16 D4-branes with gauge
group $U(4)$, with agreement with general result given in (\ref{d})

To sum up, we have seen, that from configuration of 16 non-BPS D9-branes
in IIA theory we can construct all BPS D-branes in IIA theory. In fact, we obtain
after appropriate tachyon condensation action, that describes 16 D-branes. 
  The  question remains, 
whether we can describe one single D-brane in this theory. Firstly, we can
go into     the Coulomb branch of the resulting action  and consider
one separate D-brane taking the other branes to infinity. Then we obtain 
action for single D-brane of codimension $2k+1$, but with additional 
factor $2^k$ in front of the action, which suggests, that this brane is a bound
state of $2^k$ D-branes. It is clear, that resulting D-brane looks like
ordinary D-brane of codimension $2k+1$, but its tension is different, so we 
cannot say, that this D-brane is elementary D-brane. We will see the 
same problem in analysing of WZ term in section (\ref{six}). 
 At present, we do not know, how we could
obtain action for single elementary  D-brane. It is possible that clue to this issue lies in more
general construction of tachyon condensation corresponding to 
the D-branes which do not coincide.
%%%%%%%%%%%%%%%%%%%%%%%%%%%%%%%%%%%%%%%%
%%%% non-BPS D-branes in Type IIA theory %%%%%%%%%%%%
%%%%%%%%%%%%%%%%%%%%%%%%%%%%%%%%%%%%%%%%
\section{Non-BPS D-branes in Type IIA theory}\label{four}

In this section we will construct non-BPS D-branes in Type IIA theory
from tachyon condensation in the system of $N$ non-BPS D9-branes,
following \cite{Horava3}, where
 tachyon configuration for construction of 
non-BPS D-brane of codimension $2k$ was proposed
in the form:
\begin{equation}\label{solHor}
T=\sum_{i=1}^{2k}\Gamma_i x^i
\end{equation}
where gamma matrices form a spinor representation of transverse space.
We will see that we can construct this non-BPS D-brane with almost any
effort, because this construction is directly related to the construction presented
in previous section. As in previous section we start with system $2^k$ non-BPS
D9-branes with gauge group $U(2^k)$ on their world-volume.

We generalise (\ref{solHor}) to the expression: 
\begin{equation}
T=\sum_{i=1}^{2k}\Gamma_iT(x^i)
\end{equation}
which has the same form as in case of BPS D-brane. In fact, the analysis
of equation of motion for tachyon is  the same in both situations (BPS and
non-BPS) in 
case of bosonic terms. Again tachyon condensation in this form
will leads to the D-brane of codimension $2k$ with $U(1)$ gauge symmetry
on its world-volume. But there is an difference in the case of fermionic terms.
Again we have the condition:
\begin{equation}\label{cc}
\Gamma^i\overline{\theta}_R\theta_L=0 ,i=1,...,2k
\end{equation}
but now we must expand the fermionic fields in the form
\begin{equation}
\theta_{L,R}=\sum_{i=1}^{2k+1}\theta_{L,R}^i\Gamma_i
\end{equation}
because we cannot discard the term proportional to $\Gamma^{2k+1}$
matrix. As in case of BPS D-brane we should make more general expansion
with higher antisymmetric combinations of gamma matrix but it can be
shown that these higher terms should vanish in order to obey previous equation.
When we put previous expansion of fermions into (\ref{cc}),
we get the same conditions as in case of BPS D-brane:
\begin{eqnarray}
\tr(\Gamma^i\Gamma^k\Gamma^l)\overline{\theta_R}^k
\theta_L^l=0 \nonumber \\
\overline{\theta_R}^i\theta_L^0=0 \nonumber \\
\overline{\theta_R}^0\theta_L^i=0 \nonumber \\
\end{eqnarray}

First equation leads to condition $\theta_L^i=0 ,i=1,...,2k+1$ the
second one to $\theta_R^i=0 , i=1,...,2k$ and the last equation to
condition $\theta_R^0=0$. It is important to stress that we have not
any condition on $\theta^{2k+1}_R$ due to the trivial identity
$\tr \Gamma_i\Gamma_{2k+1}=0 ,i=1,...,2k$. We than see that
we have two dynamical fermionic fields $\theta_R,\theta_L$, which is a appropriate
number of fermionic degrees of freedom  for non-BPS D-brane.
 In fact, there is also one additional tachyon
mode related to the matrix $\Gamma^{2k+1}$, which is again free dynamical
field. From that reason we must put into $F$ function in the (\ref{action2})
the form of tachyon field
\begin{equation}\label{NT}
T=T_{cs}+\mathcal{T}(y)\Gamma^{2k+1}
\end{equation}
where $T_{cs}$ is a classical solution of equation of motion, which explicitly
form was given above and $\mathcal{T}(y)$ is a free tachyonic field, which is localised on
world-volume of resulting D-brane. In this expression we implicitly presume that
$\mathcal{T}(y)$ should be multiplied with some form factor that express the fact that 
this field is localised on world-volume of the vortex. This form factor will be
the same as in case of BPS D-brane. We will write the form of
this form factor in the end of calculation.

To make thinks more clear we write  the action for system of N non-BPS D-branes
\begin{eqnarray}\label{actNT}
S=-C_9\int d^{10}x\left\{1+\frac{(2\pi\alpha')^2}{4} (\tr F_{MN}F^{MN}+2i \tr \theta_L\Gamma^MD_M\theta_L+\right. \nonumber \\
\left. +2i \tr\theta_R\Gamma^MD_M\theta_R)\right\}F(T,DT,...) \nonumber \\
\end{eqnarray}
and the form of $F$ function, which is present in (\ref{actNT}):
\begin{equation}\label{NTF}
F(T,DT..)=\frac{1}{2^k}\left [\tr D_MTD^MT+
\tr (f(T)\overline{\theta_R} \theta_L) +V(T)\right]
\end{equation}

When we put (\ref{NT}) into (\ref{NTF}) we get
\begin{equation}
F=\left[ \left\{\sum_{i=1}^{2k}\left(\frac{dT_i}{dx^i}\right)^2-
m^2T^2_i+\lambda T_i^4+\frac{m^2}{4\lambda}\right\}+
\left\{\partial_{\mu}\mathcal{T}\partial^{\mu}\mathcal{T}-
m^2\mathcal{T}^2+\lambda\mathcal{T}^4 +f(\mathcal{T})
\overline{\theta}_R\theta_L\right\}\right]
\end{equation}
where the second bracket is localised in the core of the vortex. In previous expressions
we have used the fact that the fields $T_i(x^i)$ are zero in the point
 $x^i=0$. In previous expression the partial derivative $\partial_{\mu}$ means
derivative with respect to the tangent coordinates of resulting
non-BPS D-brane. In deriving the interaction terms between fermions
and tachyon we have also used the fact that
\begin{equation}
f(\Gamma^{2k+1}\mathcal{T}) 
=\Gamma^{2k+1}f(\mathcal{T})
\end{equation}
because $f(T)$ is odd function of its argument. With using previous
equation we have the final result for interaction term between fermions
and tachyon
\begin{equation}
\tr f(T)\overline{\theta}_R\theta_L \Rightarrow
\tr (\Gamma^{2k+1}\Gamma^{2k+1})f(\mathcal{T})\overline{\theta}_R
\theta_L=2^kf(\mathcal{T})\overline{\theta}_R\theta_L
\end{equation}

Next calculation is the same as in previous 
section. With using the fact that all fields in first bracket in (\ref{actNT})
are independent on $x^i$, we can easily make integration over these coordinates
(again with appropriate insertions of convergence factor in the second bracket in
(\ref{NTF})) and we obtain the action for non-BPS Dp-brane of codimension
$2k$:
\begin{eqnarray}
S=-2^k(0.606)^{2k}C_p\int_{R^{1,p}}
d^{p+1}x \left\{1+\frac{(2\pi\alpha')^2}{4}\left[
F_{\mu\nu}F^{\mu\nu}+2\partial_{\mu}X^i\partial^{\mu}X^i+\right.\right. \nonumber \\
\left.\left. +\overline{\theta}_R\Gamma^{\mu}\partial_{\mu}\theta_R+
\overline{\theta}_L\Gamma^{\mu}\partial_{\mu}\theta_L\right]\times \right.\nonumber \\
\left.\times \left[ \partial_{\mu}\mathcal{T}\partial^{\mu}\mathcal{T}+
\left(-m^2\mathcal{T}^2+\lambda\mathcal{T}^4+\frac{m^4}{4\lambda}
\right)+f(\mathcal{T})\overline{\theta}_R\theta_L \right]\right\}\nonumber \\
\end{eqnarray} 
where $p=9-2k$. As in previous section, we have obtained the fields describing transverse
fluctuations from the  term
\begin{equation}
F_{\mu i}F^{\mu i}=2\partial_{\mu}A^i\partial^{\mu}A^i=
2\partial_{\mu}X^i\partial^{\mu}X^i
\end{equation}

We can also obtain the action for non-BPS D-brane of codimension $2k$ from the $2^k$ 
non-BPS D9-branes via tachyon condensation in the "step by step" construction presented
in section (\ref{two}) as follows:

{\bf D7-brane}: This brane has codimension $2k=2$ so that appropriate number of D9-branes
is $2$. Then we get following sequence:
\begin{equation}
2 D9\rightarrow D8+\overline{D8} \rightarrow (2) D7
\end{equation}
where the number in the bracket $(2)$ expresses the presence of 
factor $2$ in front of the action. The meaning of this number has been
discussed in previous section.

{\bf D5-brane}: It is brane of codimension $2k=4$ so that appropriate number of D9-branes
is $2^k=4$. The sequence of D-branes has a form:
\begin{equation}
4 D9\rightarrow 2 D8+2 \overline{D8}\rightarrow (2) 2 D7
\rightarrow (2)( D6+\overline{D6})\rightarrow (4) D5
\end{equation}
For non-BPS D3 and D1-brane the situation is similar.

In previous three sections we have seen that we are able to obtain action (more precisely,
kinetic part of the action) for all BPS and non-BPS D-branes in Type IIA theory. In the next
section we will discuss the emergence of D-branes in Type IIB theory.

%%%%%%%%%%%%%%%%%%%%%%%%%%%%%%%%%%%%%%%%%%%%%%%%%%%%%%%%%%%%%%
%%%%%%%%%%  D  B R A N E S      I N     T Y P E   II B  T H E O R Y %%%%%%%%%%%%%%%%%
%%%%%%%%%%%%%%%%%%%%%%%%%%%%%%%%%%%%%%%%%%%%%%%%%%%%%%%%%%%%%%
\section{D-branes in Type IIB theory}\label{five}

In this section we will discuss the emergence of BPS and non-BPS D-branes
in Type IIB theory. We start with construction of BPS D-branes.

In \cite{witen} it was proposed that all BPS D-branes in
Type IIB theory can arise as a tachyon topological solution in world-volume
of space-time filling system of D9-branes and D9-antibranes. In \cite{Kluson} 
 we have proposed the action for system of $2^{k-1}$ branes and antibranes in
the form:
 \begin{eqnarray}\label{actionB}
S=-C_9\int_{R^{1,9}}d^{10}x\left[1+\frac{(2\pi\alpha')^2}{4}\left\{\tr\left( F_{MN}F^{MN}
+2i\overline{\theta}\Gamma^{M}D_M\theta \right)\right.\right. \nonumber \\
\left.\left. +4i\tr\overline{\mathcal{B}}\Gamma^M\tilde{D}_M\mathcal{B}+ \tr\left( F'_{MN}F'^{MN}
+2i\overline{\theta}'\Gamma^MD'_M\theta'\right)\right\}\right]\times \nonumber \\
\times \frac{1}{2^k}\left\{2^k\frac{m^4}{4\lambda} + 2\tr(\tilde{D}T^{\mu}\overline{\tilde{D}_{\mu}T}
+(-m^2\tr(T\overline{T})+\lambda \tr(T\overline{T})^2)\right. 
\nonumber \\
\left. +\tr(f(T)\overline{B}\theta)+\tr_k(f(\overline{T})\overline{C}\theta')\right\} \nonumber\\
\end{eqnarray}
where $F \in U(2^{k-1}) $ is gauge field living on $2^{k-1}$ D9-branes, $F'\in U(2^{k-1})$ is gauge field
living on $2^{k-1}$ D9-antibranes , $\theta, \theta' $ are corresponding superpartners
and  tachyon $T$ and fermionic field  $\mathcal{B}$  transform in $ \bf (2^{k-1},\overline{2^{k-1}}) $
of gauge group $U(2^{k-1})\times U(2^{k-1})$.

According to ref.\cite{witen} BPS D-brane of codimension $2k$ arise from tachyon condensation
on system of space-time filling $2^{k-1}$ D9-branes and $2^{k-1}$ D9-antibranes.  We can ask the question
whether tachyon condensation in (\ref{actionB}) leads to the action of BPS D-branes. We will show
on example of step by step construction that this is really true. 

{\bf D7-brane} This is a brane of codimension $2k=2$. Than sequence of tachyon condensation has
a form:
\begin{equation}
D9+\overline{D9}\rightarrow (2){D8}\rightarrow (2)D7
\end{equation}
where the factor $(2)$ in the second term in previous expression has the same meaning as
in section (\ref{four}).

{\bf D5-brane} This is a brane of codimension $2k=4$ and tachyon condensation has a form:
\begin{equation}
2 D9+2 \overline{D9}\rightarrow 2 (2) D8\rightarrow (2)(D7+ \overline{D7}) \rightarrow
(4) D6 \rightarrow (4) D5
\end{equation}

{\bf D3-brane} This is a brane of codimension $2k=6$ and tachyon condensation has a form:
\begin{equation}
4 D9+ 4\overline{D9} \rightarrow (2)4 D8\rightarrow (2)(2D7+ 2\overline{D7})\rightarrow
(4) 2D6\rightarrow (4) (D5+\overline{D5})\rightarrow (8) D4\rightarrow (8) D3
\end{equation}

{\bf D1-brane} This is a brane of codimension $2k=8$ and tachyon condensation has a form:
\begin{eqnarray}
8 D9+ 8\overline{D9} \rightarrow (2) 8 D8\rightarrow (2) (4 D7+  4\overline{D7})\rightarrow
\nonumber \\
(4) 4D6 \rightarrow (4) (2D5+  2\overline{D5})\rightarrow (8) 2D4\rightarrow 
(8)(D3+ \overline{D3}) \rightarrow (16) D2\rightarrow (16)D1
\nonumber \\
\end{eqnarray}

We can also used the direct tachyon condensation presented in section (\ref{three}). Let us
consider the BPS D-brane of codimension $2k$. This correspond to tachyon condensation
in world-volume of $2^{k-1}$ D9-branes and D9-antibranes. As a first step we will make the tachyon
condensation in the form of kink solution, which leads to the $2^{k-1}$ non-BPS D8-branes
with gauge group $U(2^{k-1})$ (with the factor $2$ in front of action). Since transverse space to
the vortex is now $2k-1$ dimensional, the number of non-BPS D8-branes, that are needed for
construction of vortex is equal to $2^{2(k-1)/2}=2^{k-1}$, which agrees with dimension of gauge
group. Through tachyon condensation on world-volume of non-BPS D8-branes, as was presented
in section (\ref{three}) we get the stable BPS D-brane of codimension $2k$ (More precisely, the
bound state of $2^k$ D-branes of codimension $2k$.

The construction presented in previous paragraph is also appropriate for construction of
non-BPS D-branes in IIB theory of codimension $2k+1$. Following \cite{Horava3} this brane
should emerge as a tachyon vortex solution in world-volume theory of $2^k$ D9-branes and
D9-antibranes. Following the procedure in previous paragraph, we can first form a kink solution
to form $2^k$ non-BPS D8-branes in Type IIB theory. Then non-BPS D-brane, which has
originally codimension $2k+1$, appears as a object of codimension $2k$ 
in world-volume theory of non-BPS D8-brane. 
As we have seen in section (\ref{four}) on example of construction
of non-BPS D-branes in Type IIA theory (they have codimension equal to $2k$), the vortex solution, 
which looks like $T \sim \sum_{i=1}^{2k}\Gamma_i x^i$ leads to unstable non-BPS D-brane. 
We than can claim that tachyon condensation in the form in
the world-volume theory of $2^k$ D9-branes and D9-antibranes leads to the non-BPS D-brane
of codimension $2k+1$ in Type IIB theory.

We have seen in this section that we can get the correct actions for BPS and non-BPS D-branes
in Type IIB theory. In the next section we will discuss the possible form of Wess-Zumino term
for non-BPS D-branes.

%%%%%%%%%%%%%%%%%%%%%%%%%%%%%%%%%%%%%%%
%%%%% WZ term %%%%%%%%%%%%%%%%%%%%%%%%%%%%%
%%%%%%%%%%%%%%%%%%%%%%%%%%%%%%%%%%%%%%%%
\section{Wess-Zumino term for non-BPS D-brane}\label{six}
In this section we will show that tachyon condensation 
in Wess-Zumino term for non-BPS D-brane leads to correct WZ term
for BPS D-brane.  We start from generalised form of 
WZ term, which is based on previous works \cite{Billo,wilkinson} and
on the works \cite{Sen,SenA}. We hope, that this term describes
correctly the coupling between non-BPS D-branes and RR forms.

We propose RR interaction for non-BPS D-branes in the form:
\begin{equation}\label{WZA}
I_{WZ}=\mu_p\int C_p\wedge \tr\{(a_1DT+a_3(DT)^3+....+b_1DT\wedge T^2+...)
\exp \left((2\pi\alpha')F\right)\}
=\mu_p\sum_{k,l}I^A_{k,l}\end{equation}
where $\mu_p=\frac{2\pi}{(4\pi^2\alpha')^{\frac{p+1}{2}}}$ and 
where 
\begin{equation}\label{klA}
I^A_{k,l}=a_{k,l}
\int C \wedge \tr(DT)^{2k+1}T^{2l}\exp \left((2\pi\alpha')F\right) 
\end{equation}
and where $a_{k,l}$ is some numerical constant of dimension $[a_{k,l}]=l_s^{-2l}$. Unfortunately
we do not know the values of these constants, but they are not important for us
in this section, because we will not carry about numerical factors in this section.

We must say few words about (\ref{WZA}). Firstly, we can have only odd  powers of $T$ 
in WZ coupling, as was explained in \cite{Sen, SenA}. Secondly, we have replaced
ordinary derivatives with covariant derivatives as in \cite{Kluson}.
 We have included higher powers of $DT$ in (\ref{WZA}), which can be 
seen as a generalisation of \cite{Billo} and 
 we have introduced  factors proportional to
$T^{2k}$ as in ref.\cite{wilkinson}. In  this section 
we will show on various examples that proposed action (\ref{WZA}) correctly
reproduces WZ term for D-branes in Type IIA theory.

First example is "step by step" construction on 16 non-BPS D9-branes 
in Type IIA theory with gauge group $U(16)$. We have seen, that this solution
leads to action for single D0-brane. In this section, we  apply this construction for
WZ term (\ref{WZA}). We take tachyon field in the form
\begin{equation}
T(x)=\left(\begin{array}{cc}T_0(x)1_{8\times 8} & T(y) \\
             \overline{T(y)} & -T_0(x) 1_{8\times 8} \\ \end{array} \right)
\end{equation}
Standard analysis leads to 
\begin{equation}
F=\left(\begin{array}{cc} F & 0 \\
                                    0 & F' \\ \end{array}\right)
\end{equation}
Then first term gives
\begin{equation}
a_{1,0}\int C \tr DT \wedge e^{(2\pi\alpha')F)}=0.41a_{1.0}
(4\pi^2\alpha')^{1/2}\int C_p\wedge (e^{(2\pi\alpha')F)}-e^{(2\pi\alpha')F')})
\end{equation}
simply from the fact that off-diagonal terms in derivation
of tachyon do not contribute. In previous expression we have used
the form of tachyon behaviour $T(x)=T_v\tanh(\frac{mx}{\sqrt{2}})$, which
we have obtained in the second section (\ref{two}). Integration of this
function gives a factor $2$. We see that the charge of resulting D-brane is
about $0.41$ of value of correct charge of D-brane. We see two sources
of discrepancy. Firstly, we do not know  the value of constant factors $a_{k,l}$ in front of
WZ term for non-BPS D-brane. Secondly, our tachyon solution is only
rough approximation. However the form of the terms, which arise from tachyon
condensation, suggest that tachyon condensation can lead to correct
result. In the following we will not carry about numerical factors in front of
the various terms. We will also write $F$ instead
of $(2\pi\alpha')F$ and we restore the factor
$2\pi\alpha'$ in the end of the calculation.

The second term gives

\begin{eqnarray}
\int C\wedge \tr(DT^3\wedge e^F)= \nonumber \\
\int C \wedge\tr\left(\begin{array}{cc} \delta(x)dx & A \\
                                                    B & -\delta(x)dx \\
\end{array}\right)\wedge\left(
\begin{array}{cc} 0 & \tilde{D}T \\
            \overline{\tilde{D}T} & 0 \\ \end{array}\right)
\wedge \left(\begin{array}{cc} 0 & \tilde{D}T \\
               \overline{\tilde{D}T} & 0 \\ \end{array} \right)
\wedge \left(\begin{array}{cc} e^F & 0 \\
             0 & e^{F'} \\ \end{array}\right)= \nonumber \\   
=\int_{R^{1,8}}C\wedge (\tr \tilde{D}T\wedge 
\overline{\tilde{D}T}e^F-\tr \overline{\tilde{D}T}\wedge
\tilde{D}Te^{F'}) \nonumber \\
\end{eqnarray}
In the same way, third term gives
\begin{eqnarray}
\int C \wedge \tr(DT^5\wedge e^F)=
\int C\wedge \tr(D_xT \wedge (D_yT)^4e^F )=\nonumber \\
=\int C\wedge \tr(\left(
\begin{array}{cc}\delta(x) & 0 \\
                      0 & -\delta(x) \end{array}\right) \wedge 
\left(\begin{array}{cc} 0 & (\tilde{D}T\wedge \overline{\tilde{D}T})^2  \\
  (\overline{\tilde{D}T}\wedge \tilde{D}T)^2 & 0 \\ \end{array} \right)
\wedge \left(\begin{array}{cc} e^F & 0 \\
                   0  & e^{F'}  \\ \end{array} \right)= \nonumber \\             
=\int_{R^{1,8}}C\wedge (
\tr (\tilde{D}T\overline{\tilde{D}T})^2\wedge e^F -\tr 
(\overline{\tilde{D}T} \wedge \tilde{D}T)^2\wedge e^{F'} ) \nonumber \\
\end{eqnarray}

Generally, we obtain the result:
\begin{eqnarray}\label{AB}
\int C\wedge \tr (DT)^ke^{F} \Rightarrow \nonumber \\
\int_{R^{1,8}} C\wedge ( 
\tr (\tilde{D}T\wedge \overline{\tilde{D}T})^
{k-1}e^F-\tr (\overline{\tilde{D}T}\wedge \tilde{D}T)^{k-1}e^{F'}) \nonumber \\
\end{eqnarray}
Next term is
\begin{equation}
\int C \wedge DT T^2 e^F
\end{equation}
In the point $x=0$, diagonal terms in $T$ are zero, so we have
\begin{equation}
T(x=0)=\left(\begin{array}{cc} 0 & T(y) \\
                     \overline{T}(y)   &   0    \\ \end{array}\right)
\Rightarrow T^2=\left(\begin{array}{cc} 
             T \overline{T} &   0   \\
                  0 &  \overline{T} T           \\ \end{array}\right)
\end{equation}
 Then we
obtain from previous equation (Only derivative with respect to
 $x$ participates in this expression, because the  other derivatives are
off-diagonal) 
\begin{eqnarray}
\int C\wedge \tr DT T^2=\int C \wedge \tr
\left( \begin{array}{cc} \delta(x)  & 0 \\
                              0 & -\delta(x) \\ \end{array}\right)
\left (\begin{array}{cc} T\overline{T}  &  0 \\
                       0 &       \overline{T}T   \\ \end{array}\right)e^F= \nonumber \\
=\int_{R^{1,8}} C\wedge ( \tr T\overline{T}\wedge e^F -
\tr \overline{T}T \wedge e^{F'} ) \nonumber \\
\end{eqnarray}     
In the same way we obtain:
\begin{eqnarray}
\int C\wedge \tr DT T^{2k} e^F \Rightarrow \nonumber \\
\int_{R^{1,8}} C \wedge\left(\tr (T\overline{T})^ke^F -
\tr (\overline{T}T)^k e^{F'} \right ) \nonumber \\
\end{eqnarray}
Next term is
\begin{equation}
 \int C \wedge \tr (DT)^3T^2\wedge e^F 
\end{equation}
which leads to 
\begin{eqnarray}
\int C \wedge \tr 
\left(\begin{array}{cc} \delta(x) &  A \\
                                 B & -\delta(x) \\ \end{array}\right)
\wedge \left( \begin{array}{cc} \tilde{D}T\wedge \overline{\tilde{D}T} & 0 \\
                        0 &  \overline{\tilde{D}T}\wedge \tilde{D}T    \\ \end{array}\right)
\wedge \left( \begin{array}{cc} T\overline{T}  & 0 \\
                                               0     & \overline{T}T \\ \end{array}\right)
e^F =\nonumber \\
=\int C \wedge (\tr \tilde{D}T\wedge \overline{\tilde{D}T}T\overline{T} e^F-
\overline{\tilde{D}T}\wedge \tilde{D}T \overline{T}T e^{F'} ) \nonumber \\ 
\end{eqnarray}
Generally, we obtain via tachyon condensation in form a kink solution:
\begin{eqnarray}\label{Ikl}
I_{k,l}^A=\int C \wedge (DT)^{2k+1}T^{2l}e^F
\Rightarrow\nonumber \\
\int_{R^{1,8}} C \wedge (\tr \tilde{D}T\wedge \overline{\tilde{D}T})^k
(T\overline{T})^l\wedge e^F-\tr \overline{\tilde{D}T}\wedge \tilde{D}T (\overline{T}T)^l
\wedge e^{F'} )=I_{k,l}^B \nonumber \\
\end{eqnarray}
and generalised  WZ term for system  of D8-branes and  D8-antibranes is 
\begin{equation}\label{WZBantiB}
I_{WZ}^B=\mu_{8}\sum_{k,l}I_{k,l}^B
\end{equation}
We can see striking similarity with result in \cite{wilkinson}
\footnote{We will write in each step the factor $\mu_p$, because in
each step we obtain factor $(\alpha')^{1/2}$. As was explained above,
we omit the other numerical factors. We also freely use the symbol
of delta function, in order to express the fact that various fields are localised
only in the core of the vortex. We have discussed the meaning this delta
function in previous sections.}.

Now we construct kink solution on the world-volume of 8-branes and
8-antibranes. This solution was given as:
\begin{equation}
T(x,y)=T_v(x)1_{8\times 8}+iT(y)\delta(x), \ T(y)^{\dag}=T(y)
\end{equation}

We have seen, that this solution gives correct kinetic term
for non-BPS D7-brane. From this analysis we know  that $F=F'$.
We start our analysis with the terms $I_{k,0}$. We will write
$ (\tilde{D}T\wedge\overline{\tilde{D}T})^k=
(\tilde{D}T\overline{\tilde{D}T})\wedge 
(\tilde{D}T  \wedge \overline {\tilde{D}T})^{k-1} $. Now we use the 
fact, that in second bracket in previous expression we do 
not have derivation with respect $x$, so we obtain $\tilde{DT}=iDT ,
\overline{\tilde{D}T}=-iDT$ so that  the second bracket is equal
to 
\[  (DT\wedge DT)^{k-1}=(DT)^{2k-2} \]
and the first bracket is equal to
\[ dT\wedge \tilde{D}\overline{T}=d(T \tilde{D}\overline{T})=-id(T DT)
\]

where we have used the fact that all massless fields as well as $T(y)$ are
independent on $x$.
For expression  $ (\overline{\tilde{D}T}\wedge \tilde{D}T)^k
=(\overline{\tilde{D}T}\wedge \tilde{D}T)\wedge
(\overline{\tilde{D}T}\wedge \tilde{D}T)^{k-1}$ we can do the same
analysis. The second bracket is equal to $(DT)^{2k-2}$
and the first bracket leads to the result
\[ \overline{\tilde{D}}T\wedge DT=-DT\wedge \overline{\tilde{D}T}=
id(TDT) \] 

Finally, we obtain
\begin{equation}
I_{k,0}=-2i\int_{R^{1,8}}C\wedge d(\tr T (DT)^{2k-1} e^F)=
-2i\int_{R^{1,7}}C\wedge \tr T DT^{2k-1}e^F 
\end{equation}
where we have made integration over $x$. Previous expression is
 a correct result (up the sign (-2i) )for non-BPS D-branes.
The same analysis can be used for general $I_{k,l}$, because the
only difference is in presence of term
$ T\overline{T}=(iT )(-iT)=T^2 $, where we have used the fact, that
in point $x=0$, $T_0$ is zero. 
So that we obtain general result:
\begin{equation}
I_{k,l}^B\Rightarrow-2i\int_{R^{1,7}}C\wedge 
\tr (DT)^{2k-1}T^{2l} e^F=(-2i)I_{k-1,1}^A
\end{equation}
The whole WZ term is a term appropriate for non-BPS D-brane, up
the factor $ (-2i)$:
\begin{equation}
I_{WZ}=(-2i)\mu_7\sum_{k,l} I_{k,l}
\end{equation}
We can again construct kink solution
on non-BPS D7-brane with gauge group $U(8)$, leading to the
action for 4 D6-branes and 4 D6-antibranes
\footnote{We can notice, that in front of WZ term is a factor $2$.
 We will see, that in front of WZ term
for D5-brane will be factor $4$, for D3-brane will be factor $8$ and
finally for D1-brane the factor $16$ will be present. The interpretation
of these factors is the same as in section (\ref{two}).}.
 Following general recipe
(\ref{AB})
\begin{equation}
I_{k-1,l}^A \Rightarrow I_{k-1,l}^B 
\Rightarrow  I_{WZ}^A \Rightarrow I_{WZ}^B
\end{equation}
We must remember, that now the gauge group is $U(4) \times U(4)$.

Further kink solution on world-volume of brane antibrane leads
to  
\begin{equation}
I_{k-1,l}^B \Rightarrow (-2i) I_{k-2,A} , I_{WZ}^B \Rightarrow
(-2i)^2 I_{WZ}^A
\end{equation}
It is important to stress, that
\begin{equation}
I_{0,l}^B \Rightarrow 0
\end{equation}
due to the fact, that $F=F' , \ T=\overline{T} $.
As a result, we obtain action for 4 non-BPS D5-branes 
with gauge group $U(4)$. 
Further kink solution leads to 2 D4-branes and 2-D4-branes
with gauge group $U(2) \times U(2) $ and 
\begin{equation}
I_{k-2,l}^A \Rightarrow I_{k-2,l}^B
\end{equation}
Further kink solution leads to  the WZ term for 2 non-BPS D3-branes
with gauge group $U(2)$ and
\begin{equation}
I_{k-2,l}^B \Rightarrow (-2i)I^A_{k-3,l} ,
\ I_{WZ}^B \Rightarrow (-2i)^3I_{k-3,l}
\end{equation}
Next step gives action for D2-brane and D2-antibrane
\begin{equation}
I_{k-3,l}^A \Rightarrow I_{k-3,l}^B
\end{equation}
Kink solution in this system gives  non-BPS D-brane. In this
step, covariant derivative for brane+antibrane system is
\[ \tilde{D}T=dT+AT-TA' \Rightarrow dT \] 
As a result, we obtain WZ term for single non-BPS D1 brane
\begin{equation}
I=(-2i)^4\mu_1\int_{R^{1,1}} C\wedge dT \sum_{l=0}T^{2l}e^F
\end{equation}
Now we consider the last tachyon condensation. It is important to stress,
that only term with $l=0$ is nonzero, because
other terms are zero in point $x=0$ due to the fact, that $T(0)=0$. Then we obtain
standard result
\begin{equation}
I_{WZ}=2^4\mu_0 \int dt C_1
\end{equation}
We see, that this is a correct coupling of 
16 D0-branes to RR one form so together
with result in section (\ref{two}) we obtain 
right action for  BPS bound state of 16 D0-branes in IIA theory.
In the next paragraph we will discuss the tachyon condensation
in the form of vortex solution given in (\ref{three}).

In this part we will consider construction of 
general Dp-brane of codimension $2k=1$ in Type IIA theory with using
gauge theory living on $2^k$ non-BPS D9-branes.
We consider situation, when tachyon condensation
leads to D-brane of codimension $2k+1$. Following general
recipe given in \cite{Horava}, this brane can be
constructed from $2^k$ non-BPS D9-branes with
gauge group $U(2^k)$. 
For kinetic term, we have obtained correct expression
in section (\ref{three}). Now we turn to the problem of tachyon
condensation in the term given in (\ref{WZA}),(\ref{klA}):
\begin{equation}\label{WZA7}
I_{WZ}=\mu_9\sum_{k,l}I^A_{k,l}\end{equation}
where
\begin{equation}\label{klA7}
I^A_{k,l}=
\int C \wedge \tr(DT)^{2k+1}T^{2l}e^F 
\end{equation}

We know that BPS D-brane of codimenson $2k+1$ in
Type IIA theory arises from tachyon condensation in the form
\begin{equation}
T(x)=\sum_{i=1}^{2k+1}\Gamma_iT(x^i)
\end{equation}

We have analysed the form of this solution in (\ref{three}). We 
know that all $T(x^i)$ must be localised in the points $x^j=0 ,j\neq i$
and that $T(x^i)$ has a form $T(x^i)=T_v\tanh(\frac{mx}{\sqrt{2}})$.
Derivation of previous function is $\frac{m}{\sqrt{2}}(1-
\tanh^2(\frac{mx}{\sqrt{2}}))$, which has the properties similar to delta 
function (more precisely, in zero slope limit $\alpha'\rightarrow 0$ is equal
to zero almost everywhere and is finite in the point $x=0$).

We immediately can see from (\ref{klA7}) that only term $I_{2k+1,0}$ contribute to 
the form of resulting D-brane. The other terms are zero either from the fact that contain
more than $2k+1$ covariant derivatives or less and than $2k+1$, so that they 
cannot form the volume form
in the transverse space, or  contain the powers of $T$, which are zero in the core of
the vortex. As a result we obtain from $I_{2k+1,0}$:
\begin{equation}
I_{2k+1,0}=\mu_9\int_{R^{1,9}} C\wedge \tr (DT)^{2k+1} e^{(2\pi\alpha')F}
\Rightarrow 2^k\mu_p \int_{R^{1,p}} C\wedge e^{(2\pi \alpha')F}
\end{equation}
up to possible numerical factor. The factor $2^k$ comes from the trace and
field strength $F$ corresponds to abelian gauge field, as in (\ref{three}). The
emergence of the factor $2^k$ again suggest that resulting configuration is 
in fact the bound state of $2^k$ D-branes of codimension $2k+1$. We have discussed
this issue in section (\ref{three}).

Of course, as in sections (\ref{three}),  we can consider 
direct construction of WZ term for BPS D-brane of codimension
$2k+1$ in the world-volume of 16 non-BPS D9-branes.
 Consider Dp-brane of codimension $2k+1$. We know
from section (\ref{three}), that
gauge field has a form
\begin{equation}\label{a}
F=\mathrm{1}\otimes\mathcal{F}
\end{equation}
where $\mathcal{F} \in U(2^{k-4})$. 
We also know, that only term with one covariant derivative $DT$
contributes to the Wess-Zumino term, which has a form:
\begin{equation}\label{b}
I=\mu_9\int C \wedge \tr (DT)^{2k+1} \exp\left((2\pi\alpha')F\right)
\end{equation}
We know that covariant derivative has a form $DT=dt(x)\otimes 1_{2^{4-k}\times
2^{4-k}}$, where $t(x)$ has the same form as in previous paragraph.
 With using the formula $\tr (A\otimes 1)(1\otimes B)=
\tr A\tr B$ we get immediately the result (up to possible numerical factor):
\begin{equation}
I_{WZ}=2^k\mu_p \int C\wedge \tr \exp ((2\pi\alpha')\mathcal{F})
\end{equation}

where $\tr $ in previous expression goes over fundamental representation
of $U(2^{4-k})$. We again see the factor $2^k$ in front of action, which
suggests that each D-brane of codimension $2k+1$ in resulting configuration
is in fact the bound state of $2^k$ D-branes. 

We can also discuss the emergence of WZ term for non-BPS D-brane of
codimension $2m$. When we use the step by step construction proposed
in section (\ref{six}) we obtain immediately the WZ term for non-BPS D-brane.
We can also start with configuration given in (\ref{three}) for construction
of non-BPS D-branes. Again only term with $I_{2k+1,l}, \ 2k+1>2m$ contribute in this 
construction. The terms with more covariant derivative are nonzero, due to the fact
that the additional covariant derivatives contain the  derivation of tachyon field $\mathcal{T}$.
We see that we have also nonzero terms with various powers
of $T^{2l}$. This is a consequence of the fact that in the core of the vortex
the unconstrained tachyon field is nonzero. With using $T(y)^{2l}=
\mathcal{T}(\Gamma_{2k+1})^{2l}={\bf 1}\mathcal{T}$, we immediately obtain the
action for non-BPS D-brane of codimension $2k$
\begin{eqnarray}
I_{k,l}=\mu_9\int C\wedge DT^{2k+1}T^{2l}
e^{(2\pi\alpha')F} \Rightarrow \nonumber \\
I_{k-m,l}=2^k\mu_p\int C\wedge (d\mathcal{T})^{2k+1-2m} 
\mathcal{T}^{2l}e^{(2\pi\alpha')F}\nonumber \\
\end{eqnarray}

For D-branes in Type IIB theory the situation is the same as in section (\ref{five}).
For BPS D-brane of codimension $2k$ we start with WZ term for system of $2^{k-1}$
D9-branes and D9-antibranes. Then we can proceed in the same way as in (\ref{six}) 
and we will finish with WZ term for BPS D-brane. Equivalently, we could construct
the unstable D8-brane with gauge group $U(2^{k-1})$ and than proceed in the 
same way as in previous paragraph. For non-BPS D-branes the situation is basically the 
same.

\section{Conclusion}\label{eight}
In previous sections 
we have seen on many examples that our approach to the problem of tachyon
condensation gives correct form of action for D-branes in Type IIA 
and Type IIB theory. Of course, more direct calculation should
be needed for confirming our result, especially interesting appears
to us approach presented in ref.\cite{sen1,sen2,SenT,Berko}. We know that
form of our action for non-BPS D-brane is rather simple approximation,
which should be supported by more direct calculation in string theory.
On the other hand, success of our approach allows  us to claim, that
even with this simple form of action we are able to obtain correct 
form of action for BPS D-branes. It is possible that in heart of the success lies
BPS property of D-branes. It would be very nice to confirm our calculation
with direct method as in ref.\cite{sen1,sen2}.

We would like also see the direct relation to the K-theory \cite{witen,Horava}. Analysis
of D-branes in K-theory is based on nontrivial gauge fields that live on non-BPS D-branes
or on system of D-branes and D-antibranes. On the other hand we have seen that
in all our situations the gauge fields are trivial. We expect that  the nontrivial behaviour
of gauge field emerges from more general form of solution  of tachyon field. We hope to return to 
this question in the future. 

It would be  interesting to study the other theories, especially Type I theory
and $M$-theory, following \cite{Horava1}.
 It would be also interesting to study tachyon condensation
in the other systems, following \cite{Lozano1,Lozano2}. And finally, it would
be interesting to study the problem of emergence of non-Abelian gauge
symmetry for system of $N$ D-branes, that arise from tachyon condensation.

{\bf Acknowledgement:} I would like to thank Zden\v{e}k Kopeck\'{y} and Richard
von Unge for conversations.


\newpage
\begin{thebibliography}{40}
\bibitem{Kluson} J. Kluso\v{n}, \emph{"D-branes from N non-BPS D9-branes
in IIA theory"},
\jhep{0002}{017}{2000}, \hepth{9910241}.
\bibitem{Sen} A. Sen, \emph{"Supersymmetric World-volume Action for
Non-BPS  D-branes"}, \hepth{9909062}.
\bibitem{Tseytlin} A. A. Tseytlin, \emph{"Born-Infeld action, supersymmetry
and string theory"}, \hepth{9908105}.
\bibitem{SenA} A. Sen, \emph {"Non-BPS states and Branes in String Theory" }, \hepth{9904207}.
\bibitem{Lerda}A. Lerda and R. Russo,  \emph {"Stable Non-BPS states in String theory: A Pedagogical
Review" }, \hepth{9905006}.
\bibitem{Schwarz} J. Schwarz,  \emph {"TASI Lectures on Non-BPS D-Branes Systems"}, \hepth{9908144}.
\bibitem{Sen1} A. Sen, \emph{"SO(32) Spinors of Type I and other solitons on Brane-Antibrane Pair"},
 \jhep{9809}{023}{1998}, \hepth{9808141}.
\bibitem{Sen2} A. Sen, \emph{"Type I D particle and its Interactions"},
\jhep{9810}{021}{1998}, \hepth{9809111}.
\bibitem{witen} E. Witten, \emph{"D-Branes and K theory"}, hep-th/9810188.
\bibitem{Horava} P. Ho\v{r}ava, \emph{"Type IIA D-Branes, K-Theory and Matrix theory"},
\hepth{9812135}.
\bibitem{Olsen} K. Olsen and R. J. Szabo, \emph{"Brane Descent Relations in K theory"},
\hepth{9904157}.
\bibitem{Olsen2} K. Olsen and R. J. Szabo, \emph{"Constructing D-Branes From K theory"},
\hepth{9907140},
\bibitem{Sen3} A. Sen, \emph{"Tachyon condensation on Brane-Antibrane system"},
\jhep{08}{012}{1998}, \hepth{9805170}.
\bibitem{Sen4} A. Sen , {\emph "Stable non-BPS states of BPS D-particles"},
\jhep{ 08}{010}{1998}.
\bibitem{Sen5} A.Sen , \emph{ "BPS D-branes on non-supersymmetric cycles"},
\jhep{12}{021}{1998}, \hepth{9812031}.
\bibitem{Horava1} P. Ho\v{r}ava, \emph{" M theory as a holographic theory"},
\prd{59}{046004}{1999}.
\bibitem{Horava3} J. A. Harvay, P. Horava and P. Kraus,
\emph{"D-Sphalerons and the Topology of String Configurations Space"},
\hepth{0001143}
\bibitem{Billo} M. Billo, B. Craps and F. Rosse, 
\emph{"Ramond-Ramond coupling of non-BPS D-branes"}, \hepth{9905157}.
\bibitem{wilkinson} C. Kennedy and A. Wilkins,
\emph{"Ramond-Ramond Coupling on Brane-Antibrane
Systems"}, \hepth{9905185}.
\bibitem{witen3}E. Witten, \emph{"Bound States of Strings and p-Branes"},
\npb{460}{1996}{335}, \hepth{9510135}.
\bibitem{taylor} W. Taylor IV., \emph{"Lectures on D-branes, Gauge Theory and M(atrices)"},
\hepth{9801192}.
\bibitem{Witten} M. Green, J. H. Schwarz and E. Witten,
\emph{"Superstring theory, Vol.2"}, Cambridge University
Press 1987.
\bibitem{taylor2} W. Taylor IV, \emph{"Adhering 0-branes 
to 6-branes and 8-branes"}, \hepth{9705116}. 
\bibitem{sen1} A. Sen,\emph{"Universality of the Tachyon Potential"},
\hepth{9911116}.
\bibitem{sen2} A. Sen and B. Zweibach,
\emph{"Tachyon Condensation in String Field Theory"},
\hepth{9912249}.
\bibitem{Lozano1} E. Bergshoeff, E. Eyras, R. Halbersma,
C. M. Hull, Y. Lozano and J. P. van der Schaar,
\emph{"Space-time-filling Branes and Strings with Sixteen
Supercharges:}, \hepth{9812224}.
\bibitem{Lozano2} L. Houart and Y. Lozano,
\emph{"S-Duality and Brane Descent Relation"},
\hepth{9911173}. 
\bibitem{SenT}N. Berkovits, A. Sen and B. Zwiebach,
\emph{"Tachyon Condensation in Superstring Theory"},
\hepth{0002211}
\bibitem{Berko}N. Berkovits,\emph{"The Tachyon Potential
In Open String Field Theory"}, \hepth{0001084}
 
\end{thebibliography}
\end{document}